\documentclass[%
 reprint,
 amsmath,amssymb,
 aps,
]{revtex4-2}

\usepackage{graphicx}
\usepackage{subfigure}
\usepackage{dcolumn}
\usepackage{bm}

\begin{document}

\preprint{APS/123-QED}

\title{$E(5)$-like emerging $\gamma$-softness in $^{82}$Kr}

\author{Chunxiao Zhou$^{1}$}
\email{zhouchunxiao567@163.com}

\author{Tao Wang$^{2}$}%
 \email{suiyueqiaoqiao@163.com}
\affiliation{%
$^{1}$College of Mathematics and Physics Science, Hunan University of Arts and Science, Changde 415000, People's Republic of China\\
$^{2}$College of physics, Tonghua Normal University, Tonghua 134000,  People's Republic of China
}%

\date{\today}

\begin{abstract}
Recently the interacting bosom model with $SU(3)$ higher-order interactions was proposed by one of the authors, wherein unexpected $\gamma$-softness can be emerged in this new model. This stimulates further discussion on the connections of the new $\gamma$-softness and the realistic $\gamma$-soft nuclei. In this paper, $E(5)$-like $\gamma$-softness arises when the $SU(3)$ fourth-order interaction $\hat{C}_{2}^{2}[SU(3)]$ is considered. And the corresponding transitional behaviors are similar to that from the $U(5)$ limit to the $O(6)$ limit in previous IBM-1, which provides a novel perspective for understanding the new model. Low-lying spectra, $B(E2)$ values between the low-lying states and quadrupole moment of the first $2_{1}^{+}$ state are investigated. Experimental data of the $E(5)$-like nucleus $^{82}$Kr are compared with the calculated theoretical results, where the calculated low-lying level energies and the associated $B(E2)$ values fit very well with the experimental data.
\end{abstract}

\maketitle


\section{INTRODUCTION}

Nuclear shapes arising from collective dynamics between nucleons provide a fundamental concept to understand various properties of atomic nuclei \cite{Bohr1975NuclearSV}. Taking into account quadrupole degrees of freedom, nuclear shapes  are characterized as a spherical harmonic vibrator \cite{PhysRev.98.212}, an axially symmetric deformed rotor \cite{PhysRev.90.717.2} (prolate or oblate), a $\gamma$-unrelated rotor \cite{PhysRev.102.788} or a $\gamma$-rigid triaxial rotor \cite{DAVYDOV1958237}. The nuclear shapes can be elegantly described in the framework of the interacting boson model (IBM) \cite{iachello1987}. For the simplest version of IBM-1, collective properties of nuclei can be well described by Hamiltonian containing up to two-body interactions. Spectra of spherical (the $U(5)$ limit), prolate (the $SU(3)$ limit), oblate (the $\overline{SU(3)}$ limit) and $\gamma$-soft (the $O(6)$ limit) can be reproduced and shape transitional behaviors between the typical collective excitation modes are also extensively investigated in this model \cite{casten2006shape,RevModPhys.82.2155,PhysRevC.68.034326,casten2006shape,PhysRevC.73.044323,Casten_2007,PhysRevC.80.014311,CASTEN2009183,RevModPhys.82.2155,PhysRevC.85.054309}. The abrupt shape transition between different paradigms is characterized as critical point symmetry (CPS) \cite{PhysRevLett.85.3580,PhysRevLett.87.052502}, which
provides a simple parameter-free analytical treatment of transitional nuclei. In particular, the $E(5)$ CPS \cite{PhysRevLett.85.3580} corresponding to the transitional nuclei between the $U(5)$ limit and the $O(6)$ limit was initially built within the Bohr-Mottelson model \cite{Bohr1975NuclearSV} with the collective potential function being taken as an infinite square well. 

It is worth noting that the Hamiltonian only containing the lower-order interactions exposes some deficiencies, for example, it can not describe the $\gamma$-rigid triaxial deformation \cite{PhysRevC.24.684}. This means that the $\gamma$-rigid triaxial shape can not be treated on an equal footing with the prolate shape in previous IBM. The introduction of higher-order interactions $[d^{\dag}d^{\dag}d^{\dag}]^{(L)}\cdot[\tilde{d}\tilde{d}\tilde{d}]^{(L)}$ can compensate this deficiency \cite{PhysRevC.29.1420}. However, the three-body interaction does not see anything to do with the two-body terms. Meanwhile, it can be noticed that $SU(3)$ symmetry-conserving higher-order interactions have been systematically investigated to remove the degeneracy of the $\gamma$ band and the $\beta$ band \cite{PhysRevC.32.1049} and the rigid quantum asymmetric rotor within the $SU(3)$ limit has been realized \cite{PhysRevC.61.041302,PhysRevC.90.044310}.  The $SU(3)$ higher-order interactions were also investigated in \cite{ROSENSTEEL1977134,DRAAYER198561,casta1988shape,PhysRevLett.57.1124,kota20203}. Furthermore higher-order interactions can also play an important role in partial dynamical symmetry \cite{LEVIATAN201193,PhysRevLett.102.112502,PhysRevC.87.021302}. Inspired by the relationships between the $\gamma$-rigid triaxial deformation and the higher-order interactions \cite{PhysRevC.24.684,PhysRevC.29.1420}, Fortunato \emph{et al}. investigate triaxiality by introducing a cubic $Q$-consistent IBM Hamiltonian \cite{PhysRevC.84.014326}. In the $SU(3)$ limit, the cubic quadrupole interaction can describe the oblate shape, which can replace the oblate description of the $\overline{SU(3)}$ limit in previous IBM-1 and creates a new evolution path from the prolate shape to the oblate shape. It opens a new door to understand the oblate shape in realistic nuclei. The analytically solvable prolate-oblate shape phase transitional description within the $SU(3)$ limit was investigated \cite{PhysRevC.85.064312}, which offers a finite-$N$ first-order shape transition. 

These novel results \cite{PhysRevC.84.014326,PhysRevC.85.064312} encourage us to understand the existing experimental phenomena from a new perspective. The interacting bosom model with $SU(3)$ higher-order interactions (SU3-IBM) was proposed by one of the authors, in which the algebraic Hamiltonian is constructed with only considering the $U(5)$ limit and the $SU(3)$ limit  \cite{Wang2022}. Various quadrupole deformations including the $\gamma$-rigid triaxiality can be described in this $SU(3)$ limit, and the combination with the $U(5)$ limit can induce the emergence of the $\gamma$-softness. The new $\gamma$-softness was found to be intimately related with realistic $\gamma$-soft nuclei. For example, the normal states of $^{110}$Cd is newly certified that can be described by the new $\gamma$-soft rotational mode \cite{Wang2022}, which is related to the spherical nucleus puzzle \cite{PhysRevC.78.044307,garrett2010robustness,RevModPhys.83.1467,PhysRevC.86.044304,PhysRevC.86.064311,heyde2016nuclear,garrett2018critical,PhysRevLett.123.142502,PhysRevC.101.044302}. 
And it is also found that this emerging $\gamma$-softness can describe the properties of $^{196}$Pt \cite{Wang231}. In addition, the SU3-IBM can give excellent explanations to the $B(E2)$ anomaly \cite{PhysRevC.94.044327,PhysRevC.96.021301,PhysRevLett.121.022502,PhysRevC.100.034302}, where the anomaly phenomenon can be described by introducing the two $SU(3)$ third-order interactions \cite{Wang2020} or more higher-order interactions \cite{Yu2022}. On the other hand, it is found that the SU3-IBM can be also used to describe the asymmetric prolate-oblate shape phase transition in the Hf-Hg region \cite{Wang232}. The belief that triaxiality results from the competition between the prolate shape and the oblates shape \cite{Wang2022} is further exemplified, which is also discussed in the Xe-Ba region recently \cite{PhysRevLett.130.052501}. 

The $\gamma$-softness discussed in the simplest SU3-IBM has one specific feature \cite{Wang2022} that the energy of the $0_{3}^{+}$ state is nearly twice the one of the $0_{2}^{+}$ state. Thus the $0_{2}^{+}$ and $0_{3}^{+}$ states can not be close to each other, which is the major drawback when fitting the $\gamma$-soft nucleus $^{196}$Pt. This deficiency stimulates further investigations on the new $\gamma$-softness. In this paper the fourth-order interaction $\hat{C}_{2}^{2}[SU(3)]$ is introduced and new curious connection is established. The transitional behaviors similar to the evolution from the $U(5)$ limit to the $O(6)$ limit is found, and $\gamma$-softness with $E(5)$ characteristic is demonstrated. Based on the SU3-IBM, the low-lying states and the $BE(2)$ values of $^{82}$Kr are investigated, and it can be seen that the calculation results fit well with the experimental data  \cite{Rajban2021}. 

 The paper is organized as follows. In Sec. II, the Hamiltonian used in our paper is given. Sec. III shows the transitional behaviors of the Hamiltonian including the excitation spectra, $B(E2)$ transition rates and the quadrupole moment. In Sec. IV, the calculation results of $^{82}$Kr are compared with other theoretical results and experimental data. Summary of the main results and conclusions are given in Sec. V.


\section{HAMILTONIAN}

The Hamiltonian supporting the new $\gamma$-softness was proposed in \cite{Wang2022}, which is composed of two parts. One is the $d$-boson number operator $\hat{n}_{d}$ of the $U(5)$ limit. Another term is composed of various symmetry-conserving interactions of the $SU(3)$ limit. In Ref. \cite{Wang2022} the $SU(3)$ invariants are the second-order Casimir operator $-\hat{C}_{2}[SU(3)]$ and the third-order Casimir operator $\hat{C}_{3}[SU(3)]$, which describe the prolate shape and the oblate shape, respectively. Thus this formalism can be also used to the investigate of the prolate-oblate shape phase transition  \cite{PhysRevC.54.2374,PhysRevLett.87.162501,PhysRevC.68.031301}. In previous research \cite{Wang2022} the energy of the $0_{3}^{+}$ state is larger than experimental result. However, it can be noticed that the $0_{3}^{+}$ state is close to the $0_{2}^{+}$ state in many nuclei, such as $^{82}$Kr \cite{Rajban2021,Nomura2022}. In order to keep the $0_{2}^{+}$ and $0_{3}^{+}$ states close together, the new fourth-order interaction $\hat{C}_{2}^{2}[SU(3)]$ needs to be added. 
And the resulting new $\gamma$-softness can present a better description for the properties of the realistic $\gamma$-soft nuclei. It should be pointed out that there is no prior reason to guide us to do so, but the results of numerical calculations covering the parameter regions finds that there indeed exists such a new phenomenon. 

Thus the Hamiltonian in this paper is

\begin{figure}[ht]
\includegraphics[width=0.9\columnwidth]{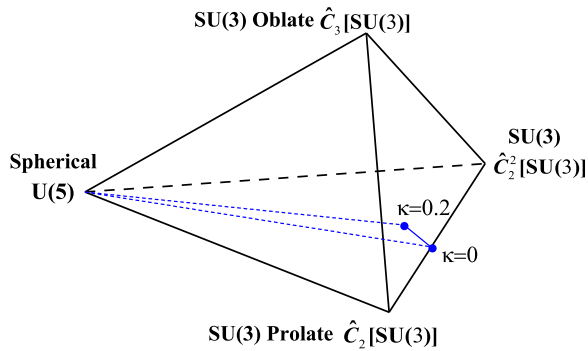}
\caption{(color online) Partial phase diagram of the SU3-IBM $\hat{H}$ used in this paper.
}\label{fig:1}
\end{figure}

\begin{widetext}
\begin{equation}\label{1}
  \hat{H}=c\left[(1-\eta)\hat{n}_{d}+\eta\left(-\frac{\hat{C}_{2}[SU(3)]}{2N}+\kappa \frac{\hat{C}_{3}[SU(3)]}{2N^{2}}+\xi\frac{\hat{C}_{2}^{2}[SU(3)]}{2N^{3}} \right)\right],
\end{equation}
\end{widetext}
where $c$ is the total fitting parameter, $0\leq\eta\leq1$, $\kappa$ and $\xi$ are the coefficient of the cubic and biquadrate interaction respectively, $N$ is the boson number. If $\eta=0$, it presents the spherical shape. If $\eta=1$, it is a combination of the $SU(3)$ limit. A schematic phase diagram of this Hamiltonian is shown in Fig. \ref{fig:1}.

The two $SU(3)$ Casimir operators are defined as
\begin{equation}\label{2}
 \hat{C}_{2}[SU(3)]=2\hat{Q}\cdot\hat{Q}+\frac{3}{4}\hat{L}\cdot\hat{L},
\end{equation}
\begin{equation}\label{3}
 \hat{C}_{3}[SU(3)]=-\frac{4\sqrt{35}}{9}[\hat{Q}\times\hat{Q}\times\hat{Q}]^{0}-\frac{\sqrt{15}}{2}[\hat{L}\times\hat{Q}\times\hat{L}]^{0}.
 \end{equation}
where $\hat{Q}$ and $\hat{L}$ are the quadrupole momentum and angular moment operators, respectively. Under the group chain $U(6)\supset SU(3)\supset O(3)$, the eigenvalues of the Casimir operators can be expressed in terms of the SU(3) irreducible representations ($\lambda$, $\mu$) as
\begin{equation}\label{5}
 \langle\hat{C}_{2}[SU(3)]\rangle=\lambda^{2}+\mu^{2}+\lambda\mu+3\lambda+3\mu,
\end{equation}
\begin{equation}\label{6}
 \langle\hat{C}_{3}[SU(3)]\rangle=\frac{1}{9}(\lambda-\mu)(2\lambda+\mu+3)(\lambda+2\mu+3).
\end{equation}

With containing higher-order interactions the Hamiltonian (\ref{1}) can generate a collective potential of a stable axially asymmetric system \cite{Rowe1985} and is relevant to anomaly phenomenon in nuclear structure \cite{Yu2022}.   

\section{ANALYSIS OF transitional behaviors}

\begin{figure}[ht]
\includegraphics[width=1\columnwidth]{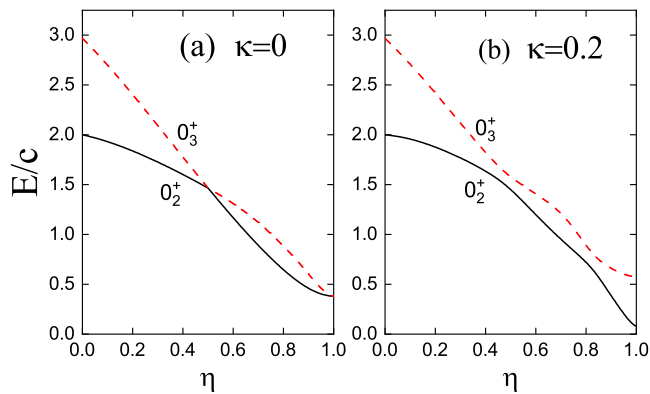}
\caption{(color online) Evolution behaviors of the $0_{2}^{+}$ and $0_{3}^{+}$ states as a function of  $\eta$ with $\xi=0.2232$ and $N=6$. (a) 
 $\kappa=0$, (b) $\kappa=0.2$.}\label{fig:2}
\end{figure}

 By including the fourth-order interaction $\hat{C}_{2}^{2}[SU(3)]$ the energy difference of the $0_{2}^{+}$ and $0_{3}^{+}$ states can be decreased. Especially, it can be found that the evolution lines of the $0_{2}^{+}$ and $0_{3}^{+}$ states can even cross with each other at $\eta=0.5$ when only considering the second-order and fourth-order interactions in Fig. \ref{fig:2} (a) , which is very similar to the evolution behaviors from the $U(5)$ limit to the $O(6)$ limit. This characteristic is a truly appealing result. The third-order interaction can dissolve the cross phenomenon and increase the energy difference of the $0_{2}^{+}$ and $0_{3}^{+}$ states, and we plot the results under the condition of $\kappa=0.2$ in Fig. \ref{fig:2} (b).
 
\begin{figure}[ht]
\subfigure{
\includegraphics[width=0.9\columnwidth]{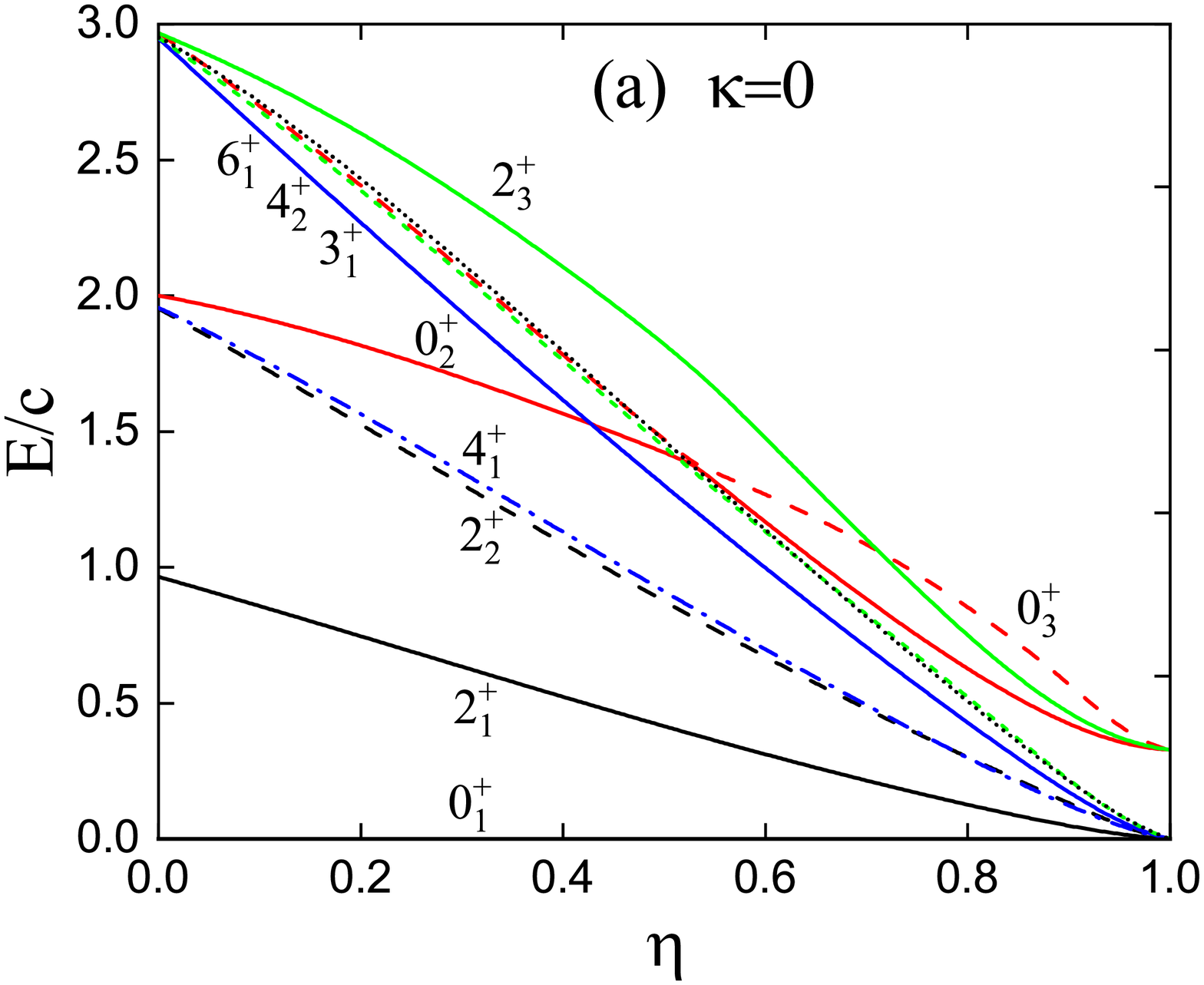}
}

\subfigure{
\includegraphics[width=0.9\columnwidth]{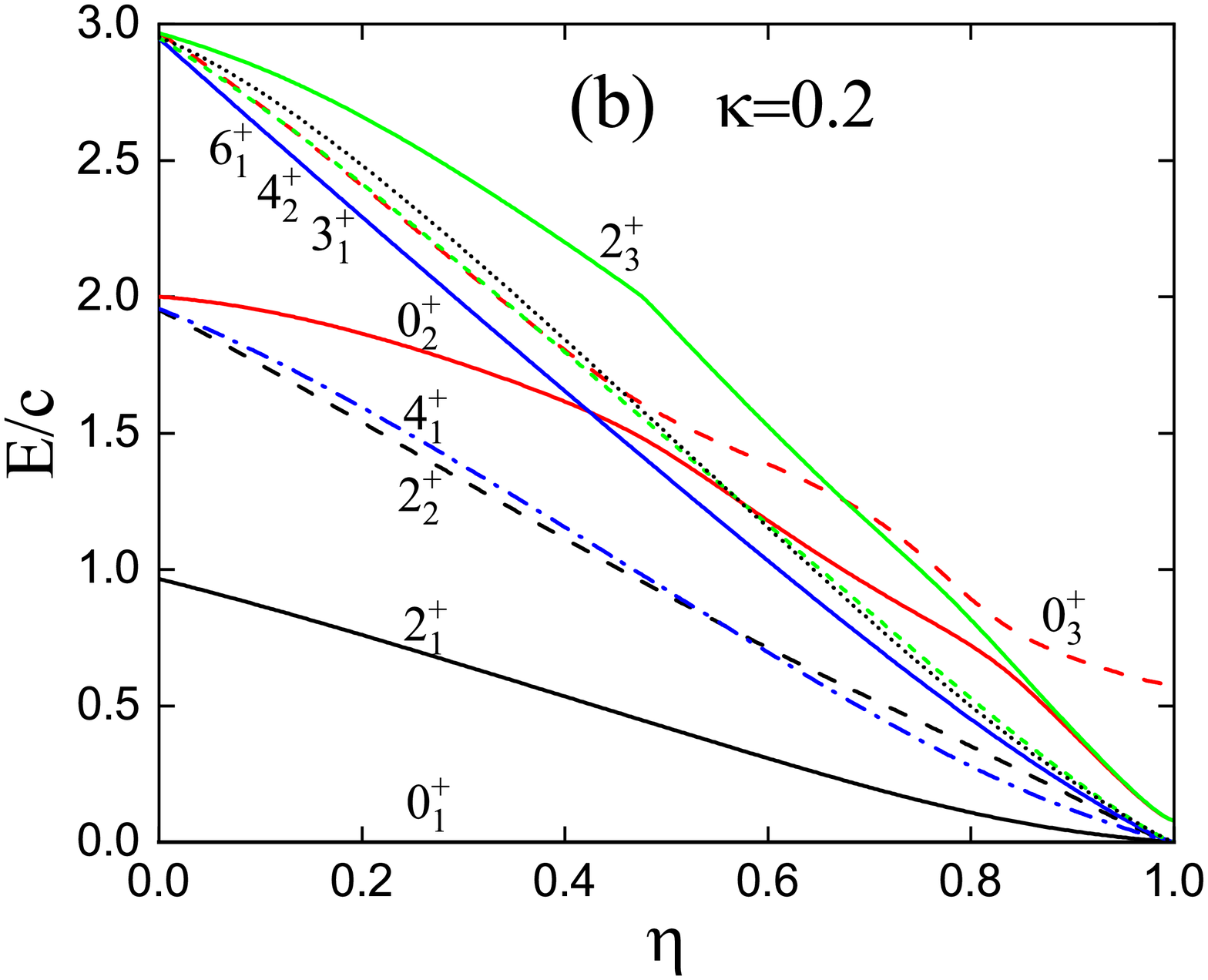}
}
\caption{(color online) Evolution behaviors of partial low-lying states as a function of $\eta$ with $\xi=0.2232$ and $N=6$. (a) $\kappa=0$, (b) $\kappa=0.2$.}\label{fig:3}
\end{figure}

The evolution behaviors of partial low-lying levels from the $U(5)$ limit to the $SU(3)$ limit as a function of $\eta$ are demonstrated in Fig. \ref{fig:3} (a), wherein only second-order and fourth-order interactions in the $SU(3)$ limit are considered. The key findings are the quasi-degeneracy of the $4_{1}^{+}$, $2_{2}^{+}$ states and the quasi-degeneracy of the $6_{1}^{+}$, $4_{2}^{+}$, $3_{1}^{+}$ states, which imply
 the $\gamma$-softness of the spectra. In addition, it can be noticed that $0_{2}^{+}$ state is higher than the quasi-degenerate $4_{1}^{+}, 2_{2}^{+}$ doublet states and crosses with the $0_{3}^{+}$ state at $\eta=0.5$. The $0_{3}^{+}$ state is degenerate with the $6_{1}^{+}$, $4_{2}^{+}$, $3_{1}^{+}$ triple states when $\eta<0.5$ and then begins to deviate from the triple states when $0.5<\eta<1$.
 As a comparison, the evolution behaviors of low-lying states with $\kappa=0.2$ are given in Fig. \ref{fig:3} (b). We can see that $0_{2}^{+}$ and $0_{3}^{+}$ states no longer intersect. However, the $\gamma$-softness is still maintained with $\kappa=0.2$, where the quasi-degenerate $4_{1}^{+}$, $2_{2}^{+}$ doublet states and the quasi-degenerate $6_{1}^{+}$, $4_{2}^{+}$, $3_{1}^{+}$ triple states are displayed in Fig. \ref{fig:3} (b). 

 The reduced transitional rate $B(E2)$ value is an important observable for the investigation of collective behaviors. Empirically definite relationships between the energy spectra and the corresponding $B(E2)$ values are expected for specific nucleus. However, such relationship can not be maintained. For example, in the Cd isotopes the energy spectra of the normal states are similar to that of the rigid spherical vibrations, while the $B(E2)$ values do not match the experimental data at all. This demonstrates that collective behaviors can not be solely determined by the energy spectra, and the corresponding $B(E2)$ values must be considered. The $E2$ operator is defined as

\begin{equation}\label{7}
\hat{T}(E2)=e\hat{Q},
\end{equation}
where $e$ is the boson effective charge. 

\begin{figure}[ht]
\includegraphics[width=1\columnwidth]{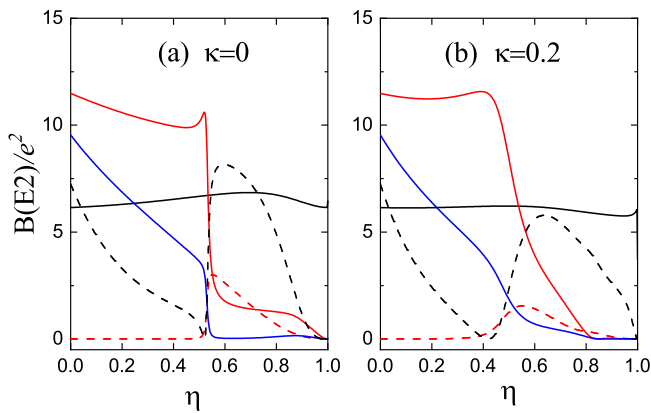}
\caption{(color online) Evolution behaviors of $B(E2;2_{1}^{+}\rightarrow 0_{1}^{+})$ (solid black line), $B(E2;0_{2}^{+}\rightarrow2_{1}^{+})$ (solid blue line), $B(E2;0_{2}^{+}\rightarrow2_{2}^{+})$ (dashed black line), $B(E2;0_{3}^{+}\rightarrow2_{1}^{+})$ (dashed red line), $B(E2;0_{3}^{+}\rightarrow2_{2}^{+})$ (solid red line) as a function of $\eta$. (a) $\kappa=0$, (b) $\kappa=0.2$. Other parameters are $N=6$ and $\xi=0.2232$.
}\label{fig:4}
\end{figure}

The evolutional behaviors of the $B(E2;2_{1}^{+}\rightarrow0_{1}^{+})$, $B(E2;0_{2}^{+}\rightarrow 2_{1}^{+})$, $B(E2;0_{2}^{+}\rightarrow 2_{2}^{+})$, $B(E2;0_{3}^{+}\rightarrow 2_{1}^{+})$, $B(E2;0_{3}^{+}\rightarrow 2_{2}^{+})$ values are plotted as functions of the parameter $\eta$ in Fig. \ref{fig:4}. In Fig. \ref{fig:4} (a), when $\kappa=0$, the slight variation is displayed on the $B(E2;2_{1}^{+}\rightarrow0_{1}^{+})$ value. However, abrupt changes are demonstrated on $B(E2;0_{2}^{+}\rightarrow2_{1}^{+})$, $B(E2;0_{2}^{+}\rightarrow2_{2}^{+})$, $B(E2;0_{3}^{+}\rightarrow2_{1}^{+})$, $B(E2;0_{3}^{+}\rightarrow2_{2}^{+})$ values just at the crossing point of the $0_{2}^{+}$ and $0_{3}^{+}$ states, which is similar to the one of the transitional behaviors from the $U(5)$ limit to the $O(6)$ limit. When $\kappa=0.2$ the curve of  $B(E2;2_{1}^{+}\rightarrow0_{1}^{+})$ value becomes flatter than the case of $\kappa=0$. In addition, the changes of $B(E2;0_{2}^{+}\rightarrow2_{1}^{+})$, $B(E2;0_{2}^{+}\rightarrow2_{2}^{+})$, $B(E2;0_{3}^{+}\rightarrow2_{1}^{+})$, $B(E2;0_{3}^{+}\rightarrow2_{2}^{+})$ values are on longer steep as those in Fig. \ref{fig:4} (a).
The variations of the $B(E2)$ values show that $\gamma$-softness emerges as the increasing of $\eta$.
 
\begin{figure}[ht]
\includegraphics[width=0.9\columnwidth]{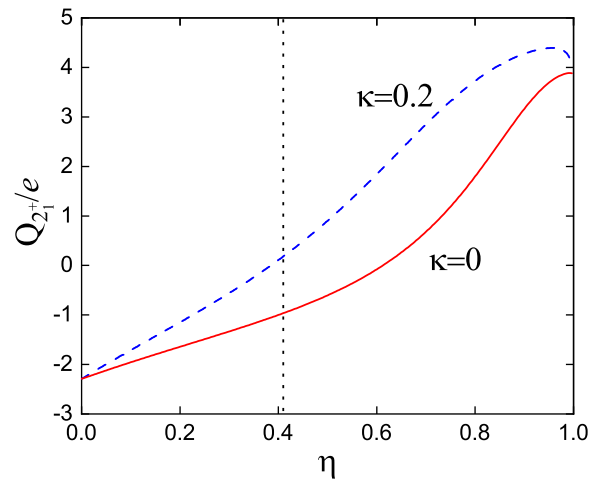}
\caption{(color online) Evolutional behaviors of the quadrupole moment of the $2_{1}^{+}$ state as a function of $\eta$ with $\kappa=0$ and $\kappa=0.2$. Other parameters are $N=6$ and $\xi=0.2232$.}\label{fig:5}
\end{figure}

If deformation is the main paradigm in nuclear structrue \cite{heyde2016nuclear}, spectroscopic quadrupole moment will be one of the most relevant quantities, especially for the prolate-oblate shape phase transition. The quadrupole moment of $2_{1}^{+}$ state is given in Fig. \ref{fig:5}. With the increasing of $\eta$ the values of quadrupole moment increase from negative to positive. This means that the shape changes from the prolate shape to the oblate shape, although they are all accompanied by a little bit deformation. In the $SU(3)$ limit, it is an oblate shape, which is induced by the fourth-order interaction $\hat{C}_{2}^{2}[SU(3)]$. It can be seen that the value of quadrupole moment with $\kappa=0.2$ is larger than the case of $\kappa=0$. This means that $\hat{C}_{3}[SU(3)]$ can make the nucleus more oblate.

\section{THEORETICAL FITTING OF $^{82}$Kr}

\begin{figure*}
\subfigure{
\includegraphics[width=2.0\columnwidth]{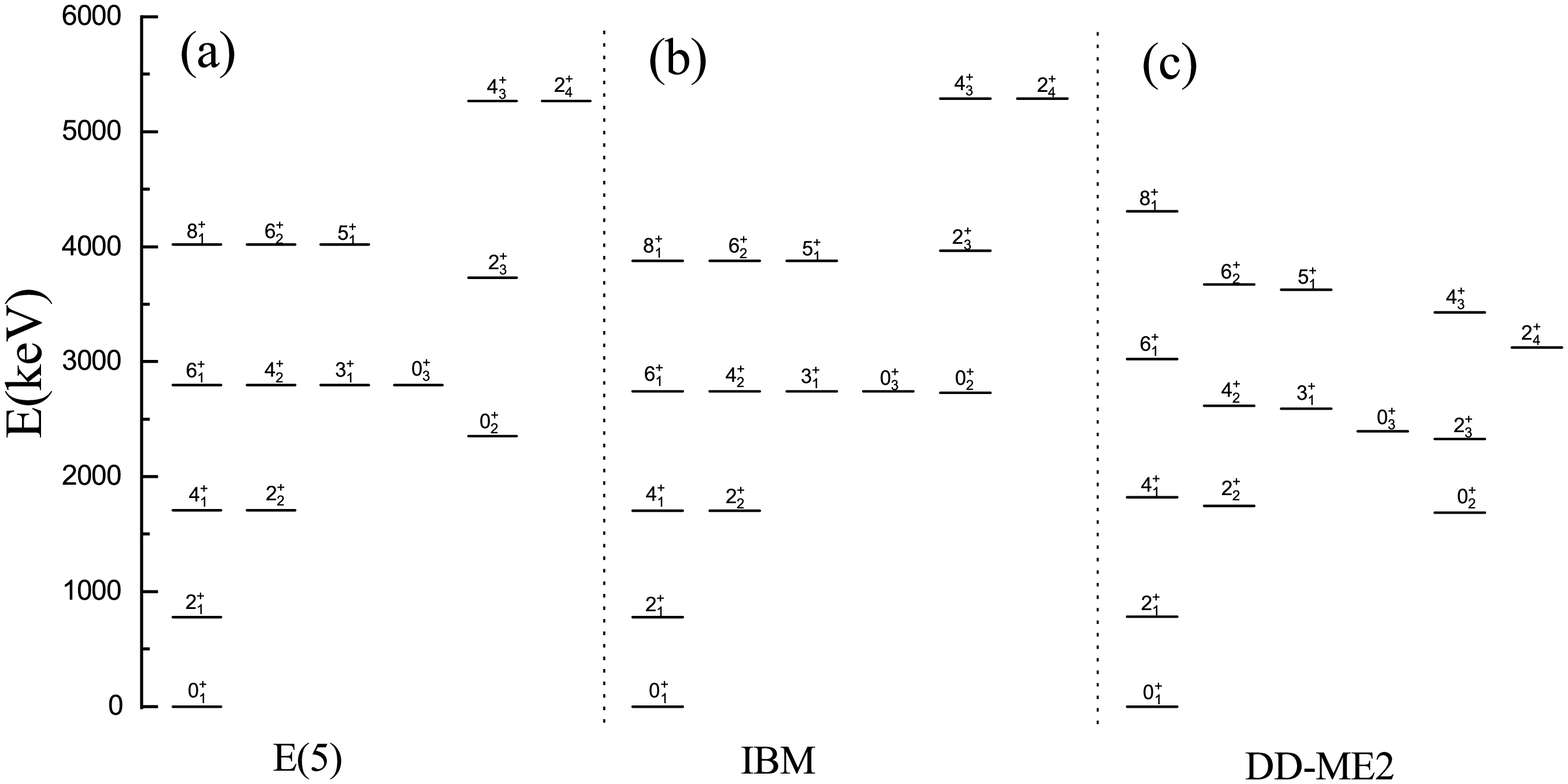}
}

\subfigure{
\includegraphics[width=2.0\columnwidth]{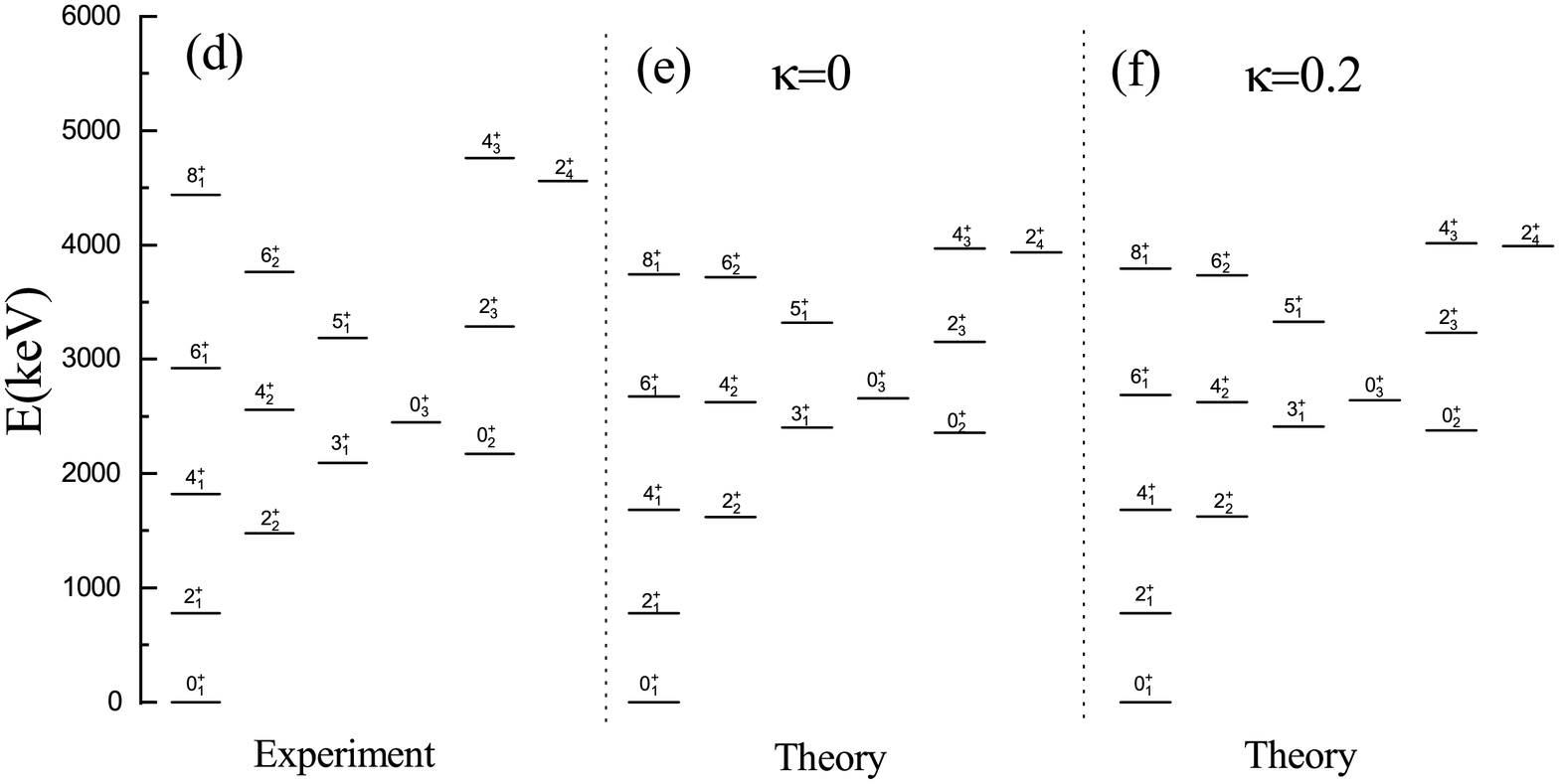}
}
\caption{Comparison of experimental and calculated partial low-lying level spectra for the $^{82}$Kr nucleus. (a), (b) and (c) are the energy spectra of $E(5)$ symmetry, IBM calculations and DD-ME2 calculations, respectively. (d) is the experimental results. (e) and (f) are the calculation results with $\kappa=0$, $c=1517.4$ KeV and $\kappa=0.2$, $c=1486.6$ KeV for $N=6$, respectively. Other parameters $\eta=0.41$, $\xi=0.2232$. The data of $E(5)$, IBM and Experiment are taken form Ref. \cite{Rajban2021}. The data of DD-ME2 are taken form Ref. \cite{Nomura2022} }\label{fig:6}
\end{figure*}

$^{82}$Kr has been identified experimentally as an empirical realization of $E(5)$ features \cite{Rajban2021,sym14102219}. The partial low-lying levels of $^{82}$Kr are demonstrated in Fig. \ref{fig:6}. The experimental values are shown in Fig. \ref{fig:6} (d) and our  calculation results are displayed in Fig. \ref{fig:6} (e) and (f) with the fitting point $\eta=0.41$. The total fitting parameters $c$ used in Fig. \ref{fig:6} (e) and (f) are determined by normalizing the calculated $2_{1}^{+}$ state to the experimental value. It can be seen that the energy levels of the $0_{2}^{+}$ and $0_{3}^{+}$ states fit well with the experiment data, and our calculated results reproduce the overall features of the experimental energy spectra. The main drawback of our theoretical calculation is that the higher levels of $8_{1}^{+}$, $4_{4}^{+}$ and $2_{5}^{+}$ are somewhat lower than the experimental results. When $\kappa=0.2$ the energy levels of $^{82}$Kr are slight higher than those of $\kappa=0$. The theoretical results are expected to be better in line with the actual results when more $SU(3)$ higher-order interactions are considered in the Hamiltonian, and this will be investigated in future. Other fitting results are also given in Fig. \ref{fig:6} for (a) $E(5)$ symmetry \cite{Rajban2021}, (b) IBM calculation \cite{Rajban2021}, and (c) DD-ME2 calculation \cite{Nomura2022}. Ref \cite{Rajban2021} shows that $^{82}$Kr can be well described by $E(5)$ symmetry. In this paper, it is shown that the properties of $^{82}$Kr can be also well described by the SU3-IBM. Thus $E(5)$-like $\gamma$-softness can emerge in the new model, which is the main conclusion of this paper.


\begin{table}
\caption{\label{tab:table1}%
Absolute $B(E2)$ values in W.u. for $E2$ transitions between the low-lying normal states in $^{82}$Kr. The data of Experimental, $E(5)$ and IBM are taken from Ref. \cite{Rajban2021}. The data of DD-ME2 are taken from Ref. \cite{Nomura2022}. The Result$^{a}$ and Result$^{b}$ are calculated with $\kappa=0$ and $\kappa=0.2$, respectively.  The corresponding effective charge of
Result$^{a}$ and Result$^{b}$ are $e$=1.8037 (W.u)$^{1/2}$ and $e$=1.8525 (W.u)$^{1/2}$, respectively.
}

\begin{ruledtabular}
\begin{tabular}{cccccccc}
\textrm{$L_{i}$}&
\textrm{$L_{f}$}&
\textrm{Expt.}&
\textrm{E(5)}&
\textrm{IBM }&
\textrm{DD-ME2}&
\textrm{Results$^{a}$}&
\textrm{Results$^{b}$}\\
\colrule
$2_{1}^{+}$ & $0_{1}^{+}$ & 21.3(10) & 21 & 21 & 23 & 21.3 & 21.3\\
$4_{1}^{+}$ & $2_{1}^{+}$ & 31.1(31) & 36 & 31 & 42 & 29.8 & 29.2\\
$2_{2}^{+}$ & $2_{1}^{+}$ & 34.6(63) & 36 & 31 & 24 & 34.6 & 35.8\\
\quad & $0_{1}^{+}$ & 1.9(2) & 0 & 0 & 0 & 0.14 & 0.45\\
$6_{1}^{+}$ & $4_{1}^{+}$ & 34.3(49) & 46 & 33 & 60  & 30.5 & 28.9\\
$4_{2}^{+}$ & $2_{2}^{+}$ & 16.8(17) & 24 & 17 & 42  & 16.8 & 17.0\\
\quad & $2_{1}^{+}$ & 0.5(1) & 0 & 0 & \quad & 0.26 & 0.44\\
\quad & $4_{1}^{+}$ & 15.8(45) & 22 & 16 & \quad & 17.6 & 18.5\\
$3_{1}^{+}$ & $2_{1}^{+}$ & 1.0(2) & 0 & 0 & 1 & 0.18 & 0.64\\
\quad & $2_{2}^{+}$ & 27.8(72) & 34 & 24 & 41 & 32.8& 32.7\\
\quad & $4_{1}^{+}$ & 9.7(75) & \quad & \quad & \quad & 15.4 & 16.1\\
$8_{1}^{+}$ & $6_{1}^{+}$ & 23.9 & 54 & 31 & 85 & 24.8 & 22.1\\
$6_{2}^{+}$ & $4_{2}^{+}$ & 21.9 & 37 & 21 & 64 & 18.5 & 17.8\\
\quad & $6_{1}^{+}$ & 7.6 & 17 & 10 & \quad & 10.1 & 10.5\\
$5_{1}^{+}$ & $3_{1}^{+}$ & 17.3(15) & 28 & 16 & 54 & 19.5 & 18.9\\
\quad & $4_{2}^{+}$ & 7.3(18) & 13 & 7 & \quad & 12.4 & 11.7\\
$0_{3}^{+}$ & $2_{1}^{+}$ & \quad & 0 & 0 & 10 & 0.02 & 0.84\\
\quad & $2_{2}^{+}$ & \quad & 46 & 33 & 44 & 32.3& 39.6\\
$0_{2}^{+}$ & $2_{1}^{+}$ & 12.1(20) & 18 & 12 & 18.0  & 15.0 & 13.0\\
\quad & $2_{2}^{+}$ & 2.0(40) & 0 & 0 &\quad & 5.0 & 0.05\\
$2_{3}^{+}$ & $0_{2}^{+}$ & 12.7(27) & 16 & 13 & 35 & 16.7& 13.2\\
\quad & $2_{2}^{+}$ & $2.5_{-1.1}^{+0.8}$ & 4 & 2 & \quad & 3.2 & 3.1\\
$4_{3}^{+}$ & $2_{3}^{+}$ & 18.5 & 26 & 21 & \quad & 21.5& 18.0\\
$2_{4}^{+}$ & $2_{3}^{+}$ & 20.5 & 26 & 21 & \quad & 21.5& 19.2\\
\end{tabular}
\end{ruledtabular}
\end{table}

The experimental and theoretical $B(E2)$ values are shown in Table \ref{tab:table1}. In the seventh and eighth columns the calculated $B(E2)$ values between some low-lying levels of $^{82}$Kr are obtained under $\kappa=0$ and $\kappa=0.2$, respectively. These results are compared with the corresponding experimental results in the third column \cite{Rajban2021}. It can be noticed that the calculated $B(E2;2_{2}^{+}\rightarrow0_{1}^{+})$ values are $0.14$ and $0.45$ for $\kappa=0$ and $\kappa=0.2$ respectively, which are smaller than the experimental result of 1.9, but are consistent with the results in $E(5)$ symmetry. Besides, significant difference are displayed on the calculated $B(E2;0_{2}^{+}\rightarrow2_{2}^{+})$ values for different $\kappa$, where the calculated results are larger and smaller than the experiment result for $\kappa=0$ and $\kappa=0.2,$ respectively. Other calculated results are fitting well with the experimental results. Thus the new theory exhibits intimate relationship with the actual properties of $^{82}$Kr and $E(5)$ symmetry.

The predicted values of the quadrupole moment of the $2_{1}^{+}$ state are -0.388 eb for $\kappa=0$ and 0.071 eb for $\kappa=0.2$, respectively (see the Fig. \ref{fig:5} at $\eta=0.41$ ). The former value means the nucleus has a prolate shape while the latter one indicates that it is a $\gamma$-soft one having a bit oblate shape. Although the energies of the low-lying levels and the $B(E2)$ values are similar to each other for $\kappa=0$ and $\kappa=0.2$, the quadrupole moments are very different.  Thus the quadrupole moment is very sensitive to the shape of the nucleus, and the experimental measurement is expected. 

\section{CONCLUSIONS}

The interacting boson model with $SU(3)$ higher-order interactions (SU3-IBM) is used to investigate the spectroscopic properties of $^{82}$Kr which was recently identified as empirical evidence for the $E(5)$ CPS. In previous studies, the $SU(3)$ limit side of the Hamiltonian in the SU3-IBM only consists of the second-order Casimir operator $\hat{C}_{2}[SU(3)]$ and the third-order Casimir operator $\hat{C}_{3}[SU(3)]$. If this Hamiltonian is used to describe the properties of the typical $\gamma$-soft nucleus $^{196}$Pt, large energy difference will exist between the $0_{2}^{+}$ and $0_{3}^{+}$ states. For overcoming this difficulty, $SU(3)$ fourth-order interaction $\hat{C}_{2}^{2}[SU(3)]$ is considered. The calculated results reveal that the energy difference of $0_{2}^{+}$ and $0_{3}^{+}$ states can be reduced even to zero in this new model. More importantly, the resulting transitional behaviors are similar to the ones from the $U(5)$ limit to the $O(6)$ limit appeared in previous IBM-1, therefore the $E(5)$-like new $\gamma$-softness is expected to be existed in the new model. The theoretical calculated energies and the $B(E2)$ values between the relevant states in $^{82}$Kr exhibit striking agreement with the experimental results and the $E(5)$ symmetry.

Based on these important findings, realistic $\gamma$-soft nuclei need to be further investigated in this new SU3-IBM, especially the traditional $O(6)$-like $\gamma$-soft nuclei, such as $^{196}$Pt. The experimental investigations have revealed that the $\gamma$-soft behaviors in $^{124-132}$Xe \cite{PhysRevC.80.061304,PhysRevC.83.044318,PhysRevC.94.024313,RAINOVSKI201011,PhysRevC.102.054304,PhysRevC.106.034311,PhysRevC.107.014324,PhysRevC.99.064321} and $^{98-102}$Zr \cite{PhysRevC.100.014319,PhysRevC.102.064314} can not be explained using the traditional $\gamma$-soft descriptions, which can be further studied in the new model. Furthermore, our results can be also used to understand the prolate-oblate shape phase transition.

\begin{acknowledgments}
The authors are grateful to K. Nomura for providing the data of DD-ME2 in Fig. \ref{fig:6} (c).
Chunxiao Zhou grateful acknowledge support from  Research Foundation of Education Bureau of Hunan
Province, China (21A0427).
\end{acknowledgments}
\nocite{*}

\bibliography{apssamp.bib}

\begin{thebibliography}{67}%
\makeatletter
\providecommand \@ifxundefined [1]{%
 \@ifx{#1\undefined}
}%
\providecommand \@ifnum [1]{%
 \ifnum #1\expandafter \@firstoftwo
 \else \expandafter \@secondoftwo
 \fi
}%
\providecommand \@ifx [1]{%
 \ifx #1\expandafter \@firstoftwo
 \else \expandafter \@secondoftwo
 \fi
}%
\providecommand \natexlab [1]{#1}%
\providecommand \enquote  [1]{``#1''}%
\providecommand \bibnamefont  [1]{#1}%
\providecommand \bibfnamefont [1]{#1}%
\providecommand \citenamefont [1]{#1}%
\providecommand \href@noop [0]{\@secondoftwo}%
\providecommand \href [0]{\begingroup \@sanitize@url \@href}%
\providecommand \@href[1]{\@@startlink{#1}\@@href}%
\providecommand \@@href[1]{\endgroup#1\@@endlink}%
\providecommand \@sanitize@url [0]{\catcode `\\12\catcode `\$12\catcode
  `\&12\catcode `\#12\catcode `\^12\catcode `\_12\catcode `\%12\relax}%
\providecommand \@@startlink[1]{}%
\providecommand \@@endlink[0]{}%
\providecommand \url  [0]{\begingroup\@sanitize@url \@url }%
\providecommand \@url [1]{\endgroup\@href {#1}{\urlprefix }}%
\providecommand \urlprefix  [0]{URL }%
\providecommand \Eprint [0]{\href }%
\providecommand \doibase [0]{https://doi.org/}%
\providecommand \selectlanguage [0]{\@gobble}%
\providecommand \bibinfo  [0]{\@secondoftwo}%
\providecommand \bibfield  [0]{\@secondoftwo}%
\providecommand \translation [1]{[#1]}%
\providecommand \BibitemOpen [0]{}%
\providecommand \bibitemStop [0]{}%
\providecommand \bibitemNoStop [0]{.\EOS\space}%
\providecommand \EOS [0]{\spacefactor3000\relax}%
\providecommand \BibitemShut  [1]{\csname bibitem#1\endcsname}%
\let\auto@bib@innerbib\@empty
\bibitem [{\citenamefont {Bohr}\ and\ \citenamefont
  {Mottelson}(1975)}]{Bohr1975NuclearSV}%
  \BibitemOpen
  \bibfield  {author} {\bibinfo {author} {\bibfnamefont {A.}~\bibnamefont
  {Bohr}}\ and\ \bibinfo {author} {\bibfnamefont {B.}~\bibnamefont
  {Mottelson}},\ }\bibfield  {title} {\bibinfo {title} {Nuclear structure,
  {Volume II}: Nuclear deformations}\ }(\bibinfo {year} {1975})\BibitemShut
  {NoStop}%
\bibitem [{\citenamefont {Scharff-Goldhaber}\ and\ \citenamefont
  {Weneser}(1955)}]{PhysRev.98.212}%
  \BibitemOpen
  \bibfield  {author} {\bibinfo {author} {\bibfnamefont {G.}~\bibnamefont
  {Scharff-Goldhaber}}\ and\ \bibinfo {author} {\bibfnamefont {J.}~\bibnamefont
  {Weneser}},\ }\bibfield  {title} {\bibinfo {title} {System of even-even
  nuclei},\ }\href {https://doi.org/10.1103/PhysRev.98.212} {\bibfield
  {journal} {\bibinfo  {journal} {Phys. Rev.}\ }\textbf {\bibinfo {volume}
  {98}},\ \bibinfo {pages} {212} (\bibinfo {year} {1955})}\BibitemShut
  {NoStop}%
\bibitem [{\citenamefont {Bohr}\ and\ \citenamefont
  {Mottelson}(1953)}]{PhysRev.90.717.2}%
  \BibitemOpen
  \bibfield  {author} {\bibinfo {author} {\bibfnamefont {A.}~\bibnamefont
  {Bohr}}\ and\ \bibinfo {author} {\bibfnamefont {B.~R.}\ \bibnamefont
  {Mottelson}},\ }\bibfield  {title} {\bibinfo {title} {Rotational states in
  even-even nuclei},\ }\href {https://doi.org/10.1103/PhysRev.90.717.2}
  {\bibfield  {journal} {\bibinfo  {journal} {Phys. Rev.}\ }\textbf {\bibinfo
  {volume} {90}},\ \bibinfo {pages} {717} (\bibinfo {year} {1953})}\BibitemShut
  {NoStop}%
\bibitem [{\citenamefont {Wilets}\ and\ \citenamefont
  {Jean}(1956)}]{PhysRev.102.788}%
  \BibitemOpen
  \bibfield  {author} {\bibinfo {author} {\bibfnamefont {L.}~\bibnamefont
  {Wilets}}\ and\ \bibinfo {author} {\bibfnamefont {M.}~\bibnamefont {Jean}},\
  }\bibfield  {title} {\bibinfo {title} {Surface oscillations in even-even
  nuclei},\ }\href {https://doi.org/10.1103/PhysRev.102.788} {\bibfield
  {journal} {\bibinfo  {journal} {Phys. Rev.}\ }\textbf {\bibinfo {volume}
  {102}},\ \bibinfo {pages} {788} (\bibinfo {year} {1956})}\BibitemShut
  {NoStop}%
\bibitem [{\citenamefont {Davydov}\ and\ \citenamefont
  {Filippov}(1958)}]{DAVYDOV1958237}%
  \BibitemOpen
  \bibfield  {author} {\bibinfo {author} {\bibfnamefont {A.}~\bibnamefont
  {Davydov}}\ and\ \bibinfo {author} {\bibfnamefont {G.}~\bibnamefont
  {Filippov}},\ }\bibfield  {title} {\bibinfo {title} {Rotational states in
  even atomic nuclei},\ }\href
  {https://doi.org/https://doi.org/10.1016/0029-5582(58)90153-6} {\bibfield
  {journal} {\bibinfo  {journal} {Nucl. Phys.}\ }\textbf {\bibinfo {volume}
  {8}},\ \bibinfo {pages} {237} (\bibinfo {year} {1958})}\BibitemShut {NoStop}%
\bibitem [{\citenamefont {Iachello}\ and\ \citenamefont
  {Arima}(1987)}]{iachello1987}%
  \BibitemOpen
  \bibfield  {author} {\bibinfo {author} {\bibfnamefont {F.}~\bibnamefont
  {Iachello}}\ and\ \bibinfo {author} {\bibfnamefont {A.}~\bibnamefont
  {Arima}},\ }\href {https://doi.org/10.1017/CBO9780511895517} {\emph {\bibinfo
  {title} {The Interacting Boson Model}}},\ Cambridge Monographs on
  Mathematical Physics\ (\bibinfo  {publisher} {Cambridge University Press},\
  \bibinfo {year} {1987})\BibitemShut {NoStop}%
\bibitem [{\citenamefont {Casten}(2006)}]{casten2006shape}%
  \BibitemOpen
  \bibfield  {author} {\bibinfo {author} {\bibfnamefont {R.}~\bibnamefont
  {Casten}},\ }\bibfield  {title} {\bibinfo {title} {Shape phase transitions
  and critical-point phenomena in atomic nuclei},\ }\href@noop {} {\bibfield
  {journal} {\bibinfo  {journal} {Nat. Phys.}\ }\textbf {\bibinfo {volume}
  {2}},\ \bibinfo {pages} {811} (\bibinfo {year} {2006})}\BibitemShut {NoStop}%
\bibitem [{\citenamefont {Cejnar}\ \emph {et~al.}(2010)\citenamefont {Cejnar},
  \citenamefont {Jolie},\ and\ \citenamefont {Casten}}]{RevModPhys.82.2155}%
  \BibitemOpen
  \bibfield  {author} {\bibinfo {author} {\bibfnamefont {P.}~\bibnamefont
  {Cejnar}}, \bibinfo {author} {\bibfnamefont {J.}~\bibnamefont {Jolie}},\ and\
  \bibinfo {author} {\bibfnamefont {R.~F.}\ \bibnamefont {Casten}},\ }\bibfield
   {title} {\bibinfo {title} {Quantum phase transitions in the shapes of atomic
  nuclei},\ }\href {https://doi.org/10.1103/RevModPhys.82.2155} {\bibfield
  {journal} {\bibinfo  {journal} {Rev. Mod. Phys.}\ }\textbf {\bibinfo {volume}
  {82}},\ \bibinfo {pages} {2155} (\bibinfo {year} {2010})}\BibitemShut
  {NoStop}%
\bibitem [{\citenamefont {Cejnar}\ \emph {et~al.}(2003)\citenamefont {Cejnar},
  \citenamefont {Heinze},\ and\ \citenamefont {Jolie}}]{PhysRevC.68.034326}%
  \BibitemOpen
  \bibfield  {author} {\bibinfo {author} {\bibfnamefont {P.}~\bibnamefont
  {Cejnar}}, \bibinfo {author} {\bibfnamefont {S.}~\bibnamefont {Heinze}},\
  and\ \bibinfo {author} {\bibfnamefont {J.}~\bibnamefont {Jolie}},\ }\bibfield
   {title} {\bibinfo {title} {Ground-state shape phase transitions in nuclei:
  Thermodynamic analogy and finite-{N} effects},\ }\href
  {https://doi.org/10.1103/PhysRevC.68.034326} {\bibfield  {journal} {\bibinfo
  {journal} {Phys. Rev. C}\ }\textbf {\bibinfo {volume} {68}},\ \bibinfo
  {pages} {034326} (\bibinfo {year} {2003})}\BibitemShut {NoStop}%
\bibitem [{\citenamefont {Luo}\ \emph {et~al.}(2006)\citenamefont {Luo},
  \citenamefont {Pan}, \citenamefont {Wang}, \citenamefont {Ning},\ and\
  \citenamefont {Draayer}}]{PhysRevC.73.044323}%
  \BibitemOpen
  \bibfield  {author} {\bibinfo {author} {\bibfnamefont {Y.~A.}\ \bibnamefont
  {Luo}}, \bibinfo {author} {\bibfnamefont {F.}~\bibnamefont {Pan}}, \bibinfo
  {author} {\bibfnamefont {T.}~\bibnamefont {Wang}}, \bibinfo {author}
  {\bibfnamefont {P.~Z.}\ \bibnamefont {Ning}},\ and\ \bibinfo {author}
  {\bibfnamefont {J.~P.}\ \bibnamefont {Draayer}},\ }\bibfield  {title}
  {\bibinfo {title} {Vibration-rotation transitional patterns in the
  $\mathit{SD}$-pair shell model},\ }\href
  {https://doi.org/10.1103/PhysRevC.73.044323} {\bibfield  {journal} {\bibinfo
  {journal} {Phys. Rev. C}\ }\textbf {\bibinfo {volume} {73}},\ \bibinfo
  {pages} {044323} (\bibinfo {year} {2006})}\BibitemShut {NoStop}%
\bibitem [{\citenamefont {Casten}\ and\ \citenamefont
  {McCutchan}(2007)}]{Casten_2007}%
  \BibitemOpen
  \bibfield  {author} {\bibinfo {author} {\bibfnamefont {R.~F.}\ \bibnamefont
  {Casten}}\ and\ \bibinfo {author} {\bibfnamefont {E.~A.}\ \bibnamefont
  {McCutchan}},\ }\bibfield  {title} {\bibinfo {title} {Quantum phase
  transitions and structural evolution in nuclei},\ }\href
  {https://doi.org/10.1088/0954-3899/34/7/R01} {\bibfield  {journal} {\bibinfo
  {journal} {J. Phys. G: Nucl. Part. Phys.}\ }\textbf {\bibinfo {volume}
  {34}},\ \bibinfo {pages} {R285} (\bibinfo {year} {2007})}\BibitemShut
  {NoStop}%
\bibitem [{\citenamefont {Luo}\ \emph {et~al.}(2009)\citenamefont {Luo},
  \citenamefont {Zhang}, \citenamefont {Meng}, \citenamefont {Pan},\ and\
  \citenamefont {Draayer}}]{PhysRevC.80.014311}%
  \BibitemOpen
  \bibfield  {author} {\bibinfo {author} {\bibfnamefont {Y.}~\bibnamefont
  {Luo}}, \bibinfo {author} {\bibfnamefont {Y.}~\bibnamefont {Zhang}}, \bibinfo
  {author} {\bibfnamefont {X.}~\bibnamefont {Meng}}, \bibinfo {author}
  {\bibfnamefont {F.}~\bibnamefont {Pan}},\ and\ \bibinfo {author}
  {\bibfnamefont {J.~P.}\ \bibnamefont {Draayer}},\ }\bibfield  {title}
  {\bibinfo {title} {Quantum phase transitional patterns in the
  $\mathit{SD}$-pair shell model},\ }\href
  {https://doi.org/10.1103/PhysRevC.80.014311} {\bibfield  {journal} {\bibinfo
  {journal} {Phys. Rev. C}\ }\textbf {\bibinfo {volume} {80}},\ \bibinfo
  {pages} {014311} (\bibinfo {year} {2009})}\BibitemShut {NoStop}%
\bibitem [{\citenamefont {Casten}(2009)}]{CASTEN2009183}%
  \BibitemOpen
  \bibfield  {author} {\bibinfo {author} {\bibfnamefont {R.}~\bibnamefont
  {Casten}},\ }\bibfield  {title} {\bibinfo {title} {Quantum phase transitions
  and structural evolution in nuclei},\ }\href
  {https://doi.org/https://doi.org/10.1016/j.ppnp.2008.06.002} {\bibfield
  {journal} {\bibinfo  {journal} {Prog. Part. Nucl. Phys.}\ }\textbf {\bibinfo
  {volume} {62}},\ \bibinfo {pages} {183} (\bibinfo {year} {2009})}\BibitemShut
  {NoStop}%
\bibitem [{\citenamefont {Kotila}\ \emph {et~al.}(2012)\citenamefont {Kotila},
  \citenamefont {Nomura}, \citenamefont {Guo}, \citenamefont {Shimizu},\ and\
  \citenamefont {Otsuka}}]{PhysRevC.85.054309}%
  \BibitemOpen
  \bibfield  {author} {\bibinfo {author} {\bibfnamefont {J.}~\bibnamefont
  {Kotila}}, \bibinfo {author} {\bibfnamefont {K.}~\bibnamefont {Nomura}},
  \bibinfo {author} {\bibfnamefont {L.}~\bibnamefont {Guo}}, \bibinfo {author}
  {\bibfnamefont {N.}~\bibnamefont {Shimizu}},\ and\ \bibinfo {author}
  {\bibfnamefont {T.}~\bibnamefont {Otsuka}},\ }\bibfield  {title} {\bibinfo
  {title} {Shape phase transitions in the interacting boson model:
  Phenomenological versus microscopic descriptions},\ }\href
  {https://doi.org/10.1103/PhysRevC.85.054309} {\bibfield  {journal} {\bibinfo
  {journal} {Phys. Rev. C}\ }\textbf {\bibinfo {volume} {85}},\ \bibinfo
  {pages} {054309} (\bibinfo {year} {2012})}\BibitemShut {NoStop}%
\bibitem [{\citenamefont {Iachello}(2000)}]{PhysRevLett.85.3580}%
  \BibitemOpen
  \bibfield  {author} {\bibinfo {author} {\bibfnamefont {F.}~\bibnamefont
  {Iachello}},\ }\bibfield  {title} {\bibinfo {title} {Dynamic symmetries at
  the critical point},\ }\href {https://doi.org/10.1103/PhysRevLett.85.3580}
  {\bibfield  {journal} {\bibinfo  {journal} {Phys. Rev. Lett.}\ }\textbf
  {\bibinfo {volume} {85}},\ \bibinfo {pages} {3580} (\bibinfo {year}
  {2000})}\BibitemShut {NoStop}%
\bibitem [{\citenamefont {Iachello}(2001)}]{PhysRevLett.87.052502}%
  \BibitemOpen
  \bibfield  {author} {\bibinfo {author} {\bibfnamefont {F.}~\bibnamefont
  {Iachello}},\ }\bibfield  {title} {\bibinfo {title} {Analytic description of
  critical point nuclei in a spherical-axially deformed shape phase
  transition},\ }\href {https://doi.org/10.1103/PhysRevLett.87.052502}
  {\bibfield  {journal} {\bibinfo  {journal} {Phys. Rev. Lett.}\ }\textbf
  {\bibinfo {volume} {87}},\ \bibinfo {pages} {052502} (\bibinfo {year}
  {2001})}\BibitemShut {NoStop}%
\bibitem [{\citenamefont {Van~Isacker}\ and\ \citenamefont
  {Chen}(1981)}]{PhysRevC.24.684}%
  \BibitemOpen
  \bibfield  {author} {\bibinfo {author} {\bibfnamefont {P.}~\bibnamefont
  {Van~Isacker}}\ and\ \bibinfo {author} {\bibfnamefont {J.-Q.}\ \bibnamefont
  {Chen}},\ }\bibfield  {title} {\bibinfo {title} {Classical limit of the
  interacting boson hamiltonian},\ }\href
  {https://doi.org/10.1103/PhysRevC.24.684} {\bibfield  {journal} {\bibinfo
  {journal} {Phys. Rev. C}\ }\textbf {\bibinfo {volume} {24}},\ \bibinfo
  {pages} {684} (\bibinfo {year} {1981})}\BibitemShut {NoStop}%
\bibitem [{\citenamefont {Heyde}\ \emph {et~al.}(1984)\citenamefont {Heyde},
  \citenamefont {Van~Isacker}, \citenamefont {Waroquier},\ and\ \citenamefont
  {Moreau}}]{PhysRevC.29.1420}%
  \BibitemOpen
  \bibfield  {author} {\bibinfo {author} {\bibfnamefont {K.}~\bibnamefont
  {Heyde}}, \bibinfo {author} {\bibfnamefont {P.}~\bibnamefont {Van~Isacker}},
  \bibinfo {author} {\bibfnamefont {M.}~\bibnamefont {Waroquier}},\ and\
  \bibinfo {author} {\bibfnamefont {J.}~\bibnamefont {Moreau}},\ }\bibfield
  {title} {\bibinfo {title} {Triaxial shapes in the interacting boson model},\
  }\href {https://doi.org/10.1103/PhysRevC.29.1420} {\bibfield  {journal}
  {\bibinfo  {journal} {Phys. Rev. C}\ }\textbf {\bibinfo {volume} {29}},\
  \bibinfo {pages} {1420} (\bibinfo {year} {1984})}\BibitemShut {NoStop}%
\bibitem [{\citenamefont {Vanden~Berghe}\ \emph {et~al.}(1985)\citenamefont
  {Vanden~Berghe}, \citenamefont {De~Meyer},\ and\ \citenamefont
  {Van~Isacker}}]{PhysRevC.32.1049}%
  \BibitemOpen
  \bibfield  {author} {\bibinfo {author} {\bibfnamefont {G.}~\bibnamefont
  {Vanden~Berghe}}, \bibinfo {author} {\bibfnamefont {H.~E.}\ \bibnamefont
  {De~Meyer}},\ and\ \bibinfo {author} {\bibfnamefont {P.}~\bibnamefont
  {Van~Isacker}},\ }\bibfield  {title} {\bibinfo {title} {Symmetry-conserving
  higher-order interaction terms in the interacting boson model},\ }\href
  {https://doi.org/10.1103/PhysRevC.32.1049} {\bibfield  {journal} {\bibinfo
  {journal} {Phys. Rev. C}\ }\textbf {\bibinfo {volume} {32}},\ \bibinfo
  {pages} {1049} (\bibinfo {year} {1985})}\BibitemShut {NoStop}%
\bibitem [{\citenamefont {Smirnov}\ \emph {et~al.}(2000)\citenamefont
  {Smirnov}, \citenamefont {Smirnova},\ and\ \citenamefont
  {Van~Isacker}}]{PhysRevC.61.041302}%
  \BibitemOpen
  \bibfield  {author} {\bibinfo {author} {\bibfnamefont {Y.~F.}\ \bibnamefont
  {Smirnov}}, \bibinfo {author} {\bibfnamefont {N.~A.}\ \bibnamefont
  {Smirnova}},\ and\ \bibinfo {author} {\bibfnamefont {P.}~\bibnamefont
  {Van~Isacker}},\ }\bibfield  {title} {\bibinfo {title} {{SU}(3) realization
  of the rigid asymmetric rotor within the interacting boson model},\ }\href
  {https://doi.org/10.1103/PhysRevC.61.041302} {\bibfield  {journal} {\bibinfo
  {journal} {Phys. Rev. C}\ }\textbf {\bibinfo {volume} {61}},\ \bibinfo
  {pages} {041302} (\bibinfo {year} {2000})}\BibitemShut {NoStop}%
\bibitem [{\citenamefont {Zhang}\ \emph {et~al.}(2014)\citenamefont {Zhang},
  \citenamefont {Pan}, \citenamefont {Dai},\ and\ \citenamefont
  {Draayer}}]{PhysRevC.90.044310}%
  \BibitemOpen
  \bibfield  {author} {\bibinfo {author} {\bibfnamefont {Y.}~\bibnamefont
  {Zhang}}, \bibinfo {author} {\bibfnamefont {F.}~\bibnamefont {Pan}}, \bibinfo
  {author} {\bibfnamefont {L.-R.}\ \bibnamefont {Dai}},\ and\ \bibinfo {author}
  {\bibfnamefont {J.~P.}\ \bibnamefont {Draayer}},\ }\bibfield  {title}
  {\bibinfo {title} {Triaxial rotor in the {SU}(3) limit of the interacting
  boson model},\ }\href {https://doi.org/10.1103/PhysRevC.90.044310} {\bibfield
   {journal} {\bibinfo  {journal} {Phys. Rev. C}\ }\textbf {\bibinfo {volume}
  {90}},\ \bibinfo {pages} {044310} (\bibinfo {year} {2014})}\BibitemShut
  {NoStop}%
\bibitem [{\citenamefont {Rosensteel}\ and\ \citenamefont
  {Rowe}(1977)}]{ROSENSTEEL1977134}%
  \BibitemOpen
  \bibfield  {author} {\bibinfo {author} {\bibfnamefont {G.}~\bibnamefont
  {Rosensteel}}\ and\ \bibinfo {author} {\bibfnamefont {D.}~\bibnamefont
  {Rowe}},\ }\bibfield  {title} {\bibinfo {title} {On the shape of deformed
  nuclei},\ }\href
  {https://doi.org/https://doi.org/10.1016/0003-4916(77)90048-3} {\bibfield
  {journal} {\bibinfo  {journal} {Ann. Phys}\ }\textbf {\bibinfo {volume}
  {104}},\ \bibinfo {pages} {134} (\bibinfo {year} {1977})}\BibitemShut
  {NoStop}%
\bibitem [{\citenamefont {Draayer}\ and\ \citenamefont
  {Rosensteel}(1985)}]{DRAAYER198561}%
  \BibitemOpen
  \bibfield  {author} {\bibinfo {author} {\bibfnamefont {J.}~\bibnamefont
  {Draayer}}\ and\ \bibinfo {author} {\bibfnamefont {G.}~\bibnamefont
  {Rosensteel}},\ }\bibfield  {title} {\bibinfo {title} {{U(3) → R(3)}
  integrity-basis spectroscopy},\ }\href
  {https://doi.org/https://doi.org/10.1016/0375-9474(85)90209-X} {\bibfield
  {journal} {\bibinfo  {journal} {Nucl. Phys. A}\ }\textbf {\bibinfo {volume}
  {439}},\ \bibinfo {pages} {61} (\bibinfo {year} {1985})}\BibitemShut
  {NoStop}%
\bibitem [{\citenamefont {Casta~nos}\ \emph {et~al.}(1988)\citenamefont
  {Casta~nos}, \citenamefont {Draayer},\ and\ \citenamefont
  {Leschber}}]{casta1988shape}%
  \BibitemOpen
  \bibfield  {author} {\bibinfo {author} {\bibfnamefont {O.}~\bibnamefont
  {Casta~nos}}, \bibinfo {author} {\bibfnamefont {J.}~\bibnamefont {Draayer}},\
  and\ \bibinfo {author} {\bibfnamefont {Y.}~\bibnamefont {Leschber}},\
  }\bibfield  {title} {\bibinfo {title} {Shape variables and the shell model},\
  }\href@noop {} {\bibfield  {journal} {\bibinfo  {journal} {Z. Phys. A Atomic
  Nuclei}\ }\textbf {\bibinfo {volume} {329}},\ \bibinfo {pages} {33} (\bibinfo
  {year} {1988})}\BibitemShut {NoStop}%
\bibitem [{\citenamefont {Elliott}\ \emph {et~al.}(1986)\citenamefont
  {Elliott}, \citenamefont {Evans},\ and\ \citenamefont
  {Van~Isacker}}]{PhysRevLett.57.1124}%
  \BibitemOpen
  \bibfield  {author} {\bibinfo {author} {\bibfnamefont {J.~P.}\ \bibnamefont
  {Elliott}}, \bibinfo {author} {\bibfnamefont {J.~A.}\ \bibnamefont {Evans}},\
  and\ \bibinfo {author} {\bibfnamefont {P.}~\bibnamefont {Van~Isacker}},\
  }\bibfield  {title} {\bibinfo {title} {Definition of the shape parameter
  $\ensuremath{\gamma}$ in the interacting-boson model},\ }\href
  {https://doi.org/10.1103/PhysRevLett.57.1124} {\bibfield  {journal} {\bibinfo
   {journal} {Phys. Rev. Lett.}\ }\textbf {\bibinfo {volume} {57}},\ \bibinfo
  {pages} {1124} (\bibinfo {year} {1986})}\BibitemShut {NoStop}%
\bibitem [{\citenamefont {Kota}(2020)}]{kota20203}%
  \BibitemOpen
  \bibfield  {author} {\bibinfo {author} {\bibfnamefont {V.}~\bibnamefont
  {Kota}},\ }\href@noop {} {\emph {\bibinfo {title} {SU (3) symmetry in atomic
  nuclei}}}\ (\bibinfo  {publisher} {Springer},\ \bibinfo {year}
  {2020})\BibitemShut {NoStop}%
\bibitem [{\citenamefont {Leviatan}(2011)}]{LEVIATAN201193}%
  \BibitemOpen
  \bibfield  {author} {\bibinfo {author} {\bibfnamefont {A.}~\bibnamefont
  {Leviatan}},\ }\bibfield  {title} {\bibinfo {title} {Partial dynamical
  symmetries},\ }\href
  {https://doi.org/https://doi.org/10.1016/j.ppnp.2010.08.001} {\bibfield
  {journal} {\bibinfo  {journal} {Prog. Part. Nucl. Phys.}\ }\textbf {\bibinfo
  {volume} {66}},\ \bibinfo {pages} {93} (\bibinfo {year} {2011})}\BibitemShut
  {NoStop}%
\bibitem [{\citenamefont {Garc\'{\i}a-Ramos}\ \emph {et~al.}(2009)\citenamefont
  {Garc\'{\i}a-Ramos}, \citenamefont {Leviatan},\ and\ \citenamefont
  {Van~Isacker}}]{PhysRevLett.102.112502}%
  \BibitemOpen
  \bibfield  {author} {\bibinfo {author} {\bibfnamefont {J.~E.}\ \bibnamefont
  {Garc\'{\i}a-Ramos}}, \bibinfo {author} {\bibfnamefont {A.}~\bibnamefont
  {Leviatan}},\ and\ \bibinfo {author} {\bibfnamefont {P.}~\bibnamefont
  {Van~Isacker}},\ }\bibfield  {title} {\bibinfo {title} {Partial dynamical
  symmetry in quantum hamiltonians with higher-order terms},\ }\href
  {https://doi.org/10.1103/PhysRevLett.102.112502} {\bibfield  {journal}
  {\bibinfo  {journal} {Phys. Rev. Lett.}\ }\textbf {\bibinfo {volume} {102}},\
  \bibinfo {pages} {112502} (\bibinfo {year} {2009})}\BibitemShut {NoStop}%
\bibitem [{\citenamefont {Leviatan}\ \emph {et~al.}(2013)\citenamefont
  {Leviatan}, \citenamefont {Garc\'{\i}a-Ramos},\ and\ \citenamefont
  {Van~Isacker}}]{PhysRevC.87.021302}%
  \BibitemOpen
  \bibfield  {author} {\bibinfo {author} {\bibfnamefont {A.}~\bibnamefont
  {Leviatan}}, \bibinfo {author} {\bibfnamefont {J.~E.}\ \bibnamefont
  {Garc\'{\i}a-Ramos}},\ and\ \bibinfo {author} {\bibfnamefont
  {P.}~\bibnamefont {Van~Isacker}},\ }\bibfield  {title} {\bibinfo {title}
  {Partial dynamical symmetry as a selection criterion for many-body
  interactions},\ }\href {https://doi.org/10.1103/PhysRevC.87.021302}
  {\bibfield  {journal} {\bibinfo  {journal} {Phys. Rev. C}\ }\textbf {\bibinfo
  {volume} {87}},\ \bibinfo {pages} {021302} (\bibinfo {year}
  {2013})}\BibitemShut {NoStop}%
\bibitem [{\citenamefont {Fortunato}\ \emph {et~al.}(2011)\citenamefont
  {Fortunato}, \citenamefont {Alonso}, \citenamefont {Arias}, \citenamefont
  {Garc\'{\i}a-Ramos},\ and\ \citenamefont {Vitturi}}]{PhysRevC.84.014326}%
  \BibitemOpen
  \bibfield  {author} {\bibinfo {author} {\bibfnamefont {L.}~\bibnamefont
  {Fortunato}}, \bibinfo {author} {\bibfnamefont {C.~E.}\ \bibnamefont
  {Alonso}}, \bibinfo {author} {\bibfnamefont {J.~M.}\ \bibnamefont {Arias}},
  \bibinfo {author} {\bibfnamefont {J.~E.}\ \bibnamefont {Garc\'{\i}a-Ramos}},\
  and\ \bibinfo {author} {\bibfnamefont {A.}~\bibnamefont {Vitturi}},\
  }\bibfield  {title} {\bibinfo {title} {Phase diagram for a cubic-$\mathrm{Q}$
  interacting boson model hamiltonian: Signs of triaxiality},\ }\href
  {https://doi.org/10.1103/PhysRevC.84.014326} {\bibfield  {journal} {\bibinfo
  {journal} {Phys. Rev. C}\ }\textbf {\bibinfo {volume} {84}},\ \bibinfo
  {pages} {014326} (\bibinfo {year} {2011})}\BibitemShut {NoStop}%
\bibitem [{\citenamefont {Zhang}\ \emph {et~al.}(2012)\citenamefont {Zhang},
  \citenamefont {Pan}, \citenamefont {Liu}, \citenamefont {Luo},\ and\
  \citenamefont {Draayer}}]{PhysRevC.85.064312}%
  \BibitemOpen
  \bibfield  {author} {\bibinfo {author} {\bibfnamefont {Y.}~\bibnamefont
  {Zhang}}, \bibinfo {author} {\bibfnamefont {F.}~\bibnamefont {Pan}}, \bibinfo
  {author} {\bibfnamefont {Y.-X.}\ \bibnamefont {Liu}}, \bibinfo {author}
  {\bibfnamefont {Y.-A.}\ \bibnamefont {Luo}},\ and\ \bibinfo {author}
  {\bibfnamefont {J.~P.}\ \bibnamefont {Draayer}},\ }\bibfield  {title}
  {\bibinfo {title} {Analytically solvable prolate-oblate shape phase
  transitional description within the {SU}(3) limit of the interacting boson
  model},\ }\href {https://doi.org/10.1103/PhysRevC.85.064312} {\bibfield
  {journal} {\bibinfo  {journal} {Phys. Rev. C}\ }\textbf {\bibinfo {volume}
  {85}},\ \bibinfo {pages} {064312} (\bibinfo {year} {2012})}\BibitemShut
  {NoStop}%
\bibitem [{\citenamefont {Wang}(2022)}]{Wang2022}%
  \BibitemOpen
  \bibfield  {author} {\bibinfo {author} {\bibfnamefont {T.}~\bibnamefont
  {Wang}},\ }\bibfield  {title} {\bibinfo {title} {New $\gamma$-soft rotation
  in the interacting boson model with {SU}(3) higher-order interactions},\
  }\href {https://doi.org/10.1088/1674-1137/ac5cb0} {\bibfield  {journal}
  {\bibinfo  {journal} {Chin. Phys. C}\ }\textbf {\bibinfo {volume} {46}},\
  \bibinfo {pages} {074101} (\bibinfo {year} {2022})}\BibitemShut {NoStop}%
\bibitem [{\citenamefont {Garrett}\ \emph {et~al.}(2008)\citenamefont
  {Garrett}, \citenamefont {Green},\ and\ \citenamefont
  {Wood}}]{PhysRevC.78.044307}%
  \BibitemOpen
  \bibfield  {author} {\bibinfo {author} {\bibfnamefont {P.~E.}\ \bibnamefont
  {Garrett}}, \bibinfo {author} {\bibfnamefont {K.~L.}\ \bibnamefont {Green}},\
  and\ \bibinfo {author} {\bibfnamefont {J.~L.}\ \bibnamefont {Wood}},\
  }\bibfield  {title} {\bibinfo {title} {Breakdown of vibrational motion in the
  isotopes $^{110\text{\ensuremath{-}}116}\mathrm{Cd}$},\ }\href
  {https://doi.org/10.1103/PhysRevC.78.044307} {\bibfield  {journal} {\bibinfo
  {journal} {Phys. Rev. C}\ }\textbf {\bibinfo {volume} {78}},\ \bibinfo
  {pages} {044307} (\bibinfo {year} {2008})}\BibitemShut {NoStop}%
\bibitem [{\citenamefont {Garrett}\ and\ \citenamefont
  {Wood}(2010)}]{garrett2010robustness}%
  \BibitemOpen
  \bibfield  {author} {\bibinfo {author} {\bibfnamefont {P.}~\bibnamefont
  {Garrett}}\ and\ \bibinfo {author} {\bibfnamefont {J.}~\bibnamefont {Wood}},\
  }\bibfield  {title} {\bibinfo {title} {On the robustness of surface
  vibrational modes: case studies in the $\mathrm{Cd}$ region},\ }\href@noop {}
  {\bibfield  {journal} {\bibinfo  {journal} {J. Phys. G: Nucl. Part. Phys.}\
  }\textbf {\bibinfo {volume} {37}},\ \bibinfo {pages} {064028} (\bibinfo
  {year} {2010})}\BibitemShut {NoStop}%
\bibitem [{\citenamefont {Heyde}\ and\ \citenamefont
  {Wood}(2011)}]{RevModPhys.83.1467}%
  \BibitemOpen
  \bibfield  {author} {\bibinfo {author} {\bibfnamefont {K.}~\bibnamefont
  {Heyde}}\ and\ \bibinfo {author} {\bibfnamefont {J.~L.}\ \bibnamefont
  {Wood}},\ }\bibfield  {title} {\bibinfo {title} {Shape coexistence in atomic
  nuclei},\ }\href {https://doi.org/10.1103/RevModPhys.83.1467} {\bibfield
  {journal} {\bibinfo  {journal} {Rev. Mod. Phys.}\ }\textbf {\bibinfo {volume}
  {83}},\ \bibinfo {pages} {1467} (\bibinfo {year} {2011})}\BibitemShut
  {NoStop}%
\bibitem [{\citenamefont {Garrett}\ \emph {et~al.}(2012)\citenamefont
  {Garrett}, \citenamefont {Bangay}, \citenamefont {Diaz~Varela}, \citenamefont
  {Ball}, \citenamefont {Cross}, \citenamefont {Demand}, \citenamefont
  {Finlay}, \citenamefont {Garnsworthy}, \citenamefont {Green}, \citenamefont
  {Hackman}, \citenamefont {Hannant}, \citenamefont {Jigmeddorj}, \citenamefont
  {Jolie}, \citenamefont {Kulp}, \citenamefont {Leach}, \citenamefont {Orce},
  \citenamefont {Phillips}, \citenamefont {Radich}, \citenamefont {Rand},
  \citenamefont {Schumaker}, \citenamefont {Svensson}, \citenamefont
  {Sumithrarachchi}, \citenamefont {Triambak}, \citenamefont {Warr},
  \citenamefont {Wong}, \citenamefont {Wood},\ and\ \citenamefont
  {Yates}}]{PhysRevC.86.044304}%
  \BibitemOpen
  \bibfield  {author} {\bibinfo {author} {\bibfnamefont {P.~E.}\ \bibnamefont
  {Garrett}}, \bibinfo {author} {\bibfnamefont {J.}~\bibnamefont {Bangay}},
  \bibinfo {author} {\bibfnamefont {A.}~\bibnamefont {Diaz~Varela}}, \bibinfo
  {author} {\bibfnamefont {G.~C.}\ \bibnamefont {Ball}}, \bibinfo {author}
  {\bibfnamefont {D.~S.}\ \bibnamefont {Cross}}, \bibinfo {author}
  {\bibfnamefont {G.~A.}\ \bibnamefont {Demand}}, \bibinfo {author}
  {\bibfnamefont {P.}~\bibnamefont {Finlay}}, \bibinfo {author} {\bibfnamefont
  {A.~B.}\ \bibnamefont {Garnsworthy}}, \bibinfo {author} {\bibfnamefont
  {K.~L.}\ \bibnamefont {Green}}, \bibinfo {author} {\bibfnamefont
  {G.}~\bibnamefont {Hackman}}, \bibinfo {author} {\bibfnamefont {C.~D.}\
  \bibnamefont {Hannant}}, \bibinfo {author} {\bibfnamefont {B.}~\bibnamefont
  {Jigmeddorj}}, \bibinfo {author} {\bibfnamefont {J.}~\bibnamefont {Jolie}},
  \bibinfo {author} {\bibfnamefont {W.~D.}\ \bibnamefont {Kulp}}, \bibinfo
  {author} {\bibfnamefont {K.~G.}\ \bibnamefont {Leach}}, \bibinfo {author}
  {\bibfnamefont {J.~N.}\ \bibnamefont {Orce}}, \bibinfo {author}
  {\bibfnamefont {A.~A.}\ \bibnamefont {Phillips}}, \bibinfo {author}
  {\bibfnamefont {A.~J.}\ \bibnamefont {Radich}}, \bibinfo {author}
  {\bibfnamefont {E.~T.}\ \bibnamefont {Rand}}, \bibinfo {author}
  {\bibfnamefont {M.~A.}\ \bibnamefont {Schumaker}}, \bibinfo {author}
  {\bibfnamefont {C.~E.}\ \bibnamefont {Svensson}}, \bibinfo {author}
  {\bibfnamefont {C.}~\bibnamefont {Sumithrarachchi}}, \bibinfo {author}
  {\bibfnamefont {S.}~\bibnamefont {Triambak}}, \bibinfo {author}
  {\bibfnamefont {N.}~\bibnamefont {Warr}}, \bibinfo {author} {\bibfnamefont
  {J.}~\bibnamefont {Wong}}, \bibinfo {author} {\bibfnamefont {J.~L.}\
  \bibnamefont {Wood}},\ and\ \bibinfo {author} {\bibfnamefont {S.~W.}\
  \bibnamefont {Yates}},\ }\bibfield  {title} {\bibinfo {title} {Detailed
  spectroscopy of ${}^{110}\mathrm{Cd}$: Evidence for weak mixing and the
  emergence of $\ensuremath{\gamma}$-soft behavior},\ }\href
  {https://doi.org/10.1103/PhysRevC.86.044304} {\bibfield  {journal} {\bibinfo
  {journal} {Phys. Rev. C}\ }\textbf {\bibinfo {volume} {86}},\ \bibinfo
  {pages} {044304} (\bibinfo {year} {2012})}\BibitemShut {NoStop}%
\bibitem [{\citenamefont {Batchelder}\ \emph {et~al.}(2012)\citenamefont
  {Batchelder}, \citenamefont {Brewer}, \citenamefont {Goans}, \citenamefont
  {Grzywacz}, \citenamefont {Griffith}, \citenamefont {Jost}, \citenamefont
  {Korgul}, \citenamefont {Liu}, \citenamefont {Paulauskas}, \citenamefont
  {Spejewski},\ and\ \citenamefont {Stracener}}]{PhysRevC.86.064311}%
  \BibitemOpen
  \bibfield  {author} {\bibinfo {author} {\bibfnamefont {J.~C.}\ \bibnamefont
  {Batchelder}}, \bibinfo {author} {\bibfnamefont {N.~T.}\ \bibnamefont
  {Brewer}}, \bibinfo {author} {\bibfnamefont {R.~E.}\ \bibnamefont {Goans}},
  \bibinfo {author} {\bibfnamefont {R.}~\bibnamefont {Grzywacz}}, \bibinfo
  {author} {\bibfnamefont {B.~O.}\ \bibnamefont {Griffith}}, \bibinfo {author}
  {\bibfnamefont {C.}~\bibnamefont {Jost}}, \bibinfo {author} {\bibfnamefont
  {A.}~\bibnamefont {Korgul}}, \bibinfo {author} {\bibfnamefont {S.~H.}\
  \bibnamefont {Liu}}, \bibinfo {author} {\bibfnamefont {S.~V.}\ \bibnamefont
  {Paulauskas}}, \bibinfo {author} {\bibfnamefont {E.~H.}\ \bibnamefont
  {Spejewski}},\ and\ \bibinfo {author} {\bibfnamefont {D.~W.}\ \bibnamefont
  {Stracener}},\ }\bibfield  {title} {\bibinfo {title} {Low-lying collective
  states in ${}^{120}\mathrm{Cd}$ populated by $\ensuremath{\beta}$ decay of
  ${}^{120}\mathrm{Ag}$: Breakdown of the anharmonic vibrator model at the
  three-phonon level},\ }\href {https://doi.org/10.1103/PhysRevC.86.064311}
  {\bibfield  {journal} {\bibinfo  {journal} {Phys. Rev. C}\ }\textbf {\bibinfo
  {volume} {86}},\ \bibinfo {pages} {064311} (\bibinfo {year}
  {2012})}\BibitemShut {NoStop}%
\bibitem [{\citenamefont {Heyde}\ and\ \citenamefont
  {Wood}(2016)}]{heyde2016nuclear}%
  \BibitemOpen
  \bibfield  {author} {\bibinfo {author} {\bibfnamefont {K.}~\bibnamefont
  {Heyde}}\ and\ \bibinfo {author} {\bibfnamefont {J.}~\bibnamefont {Wood}},\
  }\bibfield  {title} {\bibinfo {title} {Nuclear shapes: from earliest ideas to
  multiple shape coexisting structures},\ }\href@noop {} {\bibfield  {journal}
  {\bibinfo  {journal} {Phys. Scr.}\ }\textbf {\bibinfo {volume} {91}},\
  \bibinfo {pages} {083008} (\bibinfo {year} {2016})}\BibitemShut {NoStop}%
\bibitem [{\citenamefont {Garrett}\ \emph {et~al.}(2018)\citenamefont
  {Garrett}, \citenamefont {Wood},\ and\ \citenamefont
  {Yates}}]{garrett2018critical}%
  \BibitemOpen
  \bibfield  {author} {\bibinfo {author} {\bibfnamefont {P.}~\bibnamefont
  {Garrett}}, \bibinfo {author} {\bibfnamefont {J.}~\bibnamefont {Wood}},\ and\
  \bibinfo {author} {\bibfnamefont {S.}~\bibnamefont {Yates}},\ }\bibfield
  {title} {\bibinfo {title} {Critical insights into nuclear collectivity from
  complementary nuclear spectroscopic methods},\ }\href@noop {} {\bibfield
  {journal} {\bibinfo  {journal} {Phys. Scr.}\ }\textbf {\bibinfo {volume}
  {93}},\ \bibinfo {pages} {063001} (\bibinfo {year} {2018})}\BibitemShut
  {NoStop}%
\bibitem [{\citenamefont {Garrett}\ \emph {et~al.}(2019)\citenamefont
  {Garrett}, \citenamefont {Rodr\'{\i}guez}, \citenamefont {Varela},
  \citenamefont {Green}, \citenamefont {Bangay}, \citenamefont {Finlay},
  \citenamefont {Austin}, \citenamefont {Ball}, \citenamefont {Bandyopadhyay},
  \citenamefont {Bildstein}, \citenamefont {Colosimo}, \citenamefont {Cross},
  \citenamefont {Demand}, \citenamefont {Finlay}, \citenamefont {Garnsworthy},
  \citenamefont {Grinyer}, \citenamefont {Hackman}, \citenamefont {Jigmeddorj},
  \citenamefont {Jolie}, \citenamefont {Kulp}, \citenamefont {Leach},
  \citenamefont {Morton}, \citenamefont {Orce}, \citenamefont {Pearson},
  \citenamefont {Phillips}, \citenamefont {Radich}, \citenamefont {Rand},
  \citenamefont {Schumaker}, \citenamefont {Svensson}, \citenamefont
  {Sumithrarachchi}, \citenamefont {Triambak}, \citenamefont {Warr},
  \citenamefont {Wong}, \citenamefont {Wood},\ and\ \citenamefont
  {Yates}}]{PhysRevLett.123.142502}%
  \BibitemOpen
  \bibfield  {author} {\bibinfo {author} {\bibfnamefont {P.~E.}\ \bibnamefont
  {Garrett}}, \bibinfo {author} {\bibfnamefont {T.~R.}\ \bibnamefont
  {Rodr\'{\i}guez}}, \bibinfo {author} {\bibfnamefont {A.~D.}\ \bibnamefont
  {Varela}}, \bibinfo {author} {\bibfnamefont {K.~L.}\ \bibnamefont {Green}},
  \bibinfo {author} {\bibfnamefont {J.}~\bibnamefont {Bangay}}, \bibinfo
  {author} {\bibfnamefont {A.}~\bibnamefont {Finlay}}, \bibinfo {author}
  {\bibfnamefont {R.~A.~E.}\ \bibnamefont {Austin}}, \bibinfo {author}
  {\bibfnamefont {G.~C.}\ \bibnamefont {Ball}}, \bibinfo {author}
  {\bibfnamefont {D.~S.}\ \bibnamefont {Bandyopadhyay}}, \bibinfo {author}
  {\bibfnamefont {V.}~\bibnamefont {Bildstein}}, \bibinfo {author}
  {\bibfnamefont {S.}~\bibnamefont {Colosimo}}, \bibinfo {author}
  {\bibfnamefont {D.~S.}\ \bibnamefont {Cross}}, \bibinfo {author}
  {\bibfnamefont {G.~A.}\ \bibnamefont {Demand}}, \bibinfo {author}
  {\bibfnamefont {P.}~\bibnamefont {Finlay}}, \bibinfo {author} {\bibfnamefont
  {A.~B.}\ \bibnamefont {Garnsworthy}}, \bibinfo {author} {\bibfnamefont
  {G.~F.}\ \bibnamefont {Grinyer}}, \bibinfo {author} {\bibfnamefont
  {G.}~\bibnamefont {Hackman}}, \bibinfo {author} {\bibfnamefont
  {B.}~\bibnamefont {Jigmeddorj}}, \bibinfo {author} {\bibfnamefont
  {J.}~\bibnamefont {Jolie}}, \bibinfo {author} {\bibfnamefont {W.~D.}\
  \bibnamefont {Kulp}}, \bibinfo {author} {\bibfnamefont {K.~G.}\ \bibnamefont
  {Leach}}, \bibinfo {author} {\bibfnamefont {A.~C.}\ \bibnamefont {Morton}},
  \bibinfo {author} {\bibfnamefont {J.~N.}\ \bibnamefont {Orce}}, \bibinfo
  {author} {\bibfnamefont {C.~J.}\ \bibnamefont {Pearson}}, \bibinfo {author}
  {\bibfnamefont {A.~A.}\ \bibnamefont {Phillips}}, \bibinfo {author}
  {\bibfnamefont {A.~J.}\ \bibnamefont {Radich}}, \bibinfo {author}
  {\bibfnamefont {E.~T.}\ \bibnamefont {Rand}}, \bibinfo {author}
  {\bibfnamefont {M.~A.}\ \bibnamefont {Schumaker}}, \bibinfo {author}
  {\bibfnamefont {C.~E.}\ \bibnamefont {Svensson}}, \bibinfo {author}
  {\bibfnamefont {C.}~\bibnamefont {Sumithrarachchi}}, \bibinfo {author}
  {\bibfnamefont {S.}~\bibnamefont {Triambak}}, \bibinfo {author}
  {\bibfnamefont {N.}~\bibnamefont {Warr}}, \bibinfo {author} {\bibfnamefont
  {J.}~\bibnamefont {Wong}}, \bibinfo {author} {\bibfnamefont {J.~L.}\
  \bibnamefont {Wood}},\ and\ \bibinfo {author} {\bibfnamefont {S.~W.}\
  \bibnamefont {Yates}},\ }\bibfield  {title} {\bibinfo {title} {Multiple shape
  coexistence in $^{110,112}\mathrm{Cd}$},\ }\href
  {https://doi.org/10.1103/PhysRevLett.123.142502} {\bibfield  {journal}
  {\bibinfo  {journal} {Phys. Rev. Lett.}\ }\textbf {\bibinfo {volume} {123}},\
  \bibinfo {pages} {142502} (\bibinfo {year} {2019})}\BibitemShut {NoStop}%
\bibitem [{\citenamefont {Garrett}\ \emph {et~al.}(2020)\citenamefont
  {Garrett}, \citenamefont {Rodr\'{\i}guez}, \citenamefont {Diaz~Varela},
  \citenamefont {Green}, \citenamefont {Bangay}, \citenamefont {Finlay},
  \citenamefont {Austin}, \citenamefont {Ball}, \citenamefont {Bandyopadhyay},
  \citenamefont {Bildstein}, \citenamefont {Colosimo}, \citenamefont {Cross},
  \citenamefont {Demand}, \citenamefont {Finlay}, \citenamefont {Garnsworthy},
  \citenamefont {Grinyer}, \citenamefont {Hackman}, \citenamefont {Jigmeddorj},
  \citenamefont {Jolie}, \citenamefont {Kulp}, \citenamefont {Leach},
  \citenamefont {Morton}, \citenamefont {Orce}, \citenamefont {Pearson},
  \citenamefont {Phillips}, \citenamefont {Radich}, \citenamefont {Rand},
  \citenamefont {Schumaker}, \citenamefont {Svensson}, \citenamefont
  {Sumithrarachchi}, \citenamefont {Triambak}, \citenamefont {Warr},
  \citenamefont {Wong}, \citenamefont {Wood},\ and\ \citenamefont
  {Yates}}]{PhysRevC.101.044302}%
  \BibitemOpen
  \bibfield  {author} {\bibinfo {author} {\bibfnamefont {P.~E.}\ \bibnamefont
  {Garrett}}, \bibinfo {author} {\bibfnamefont {T.~R.}\ \bibnamefont
  {Rodr\'{\i}guez}}, \bibinfo {author} {\bibfnamefont {A.}~\bibnamefont
  {Diaz~Varela}}, \bibinfo {author} {\bibfnamefont {K.~L.}\ \bibnamefont
  {Green}}, \bibinfo {author} {\bibfnamefont {J.}~\bibnamefont {Bangay}},
  \bibinfo {author} {\bibfnamefont {A.}~\bibnamefont {Finlay}}, \bibinfo
  {author} {\bibfnamefont {R.~A.~E.}\ \bibnamefont {Austin}}, \bibinfo {author}
  {\bibfnamefont {G.~C.}\ \bibnamefont {Ball}}, \bibinfo {author}
  {\bibfnamefont {D.~S.}\ \bibnamefont {Bandyopadhyay}}, \bibinfo {author}
  {\bibfnamefont {V.}~\bibnamefont {Bildstein}}, \bibinfo {author}
  {\bibfnamefont {S.}~\bibnamefont {Colosimo}}, \bibinfo {author}
  {\bibfnamefont {D.~S.}\ \bibnamefont {Cross}}, \bibinfo {author}
  {\bibfnamefont {G.~A.}\ \bibnamefont {Demand}}, \bibinfo {author}
  {\bibfnamefont {P.}~\bibnamefont {Finlay}}, \bibinfo {author} {\bibfnamefont
  {A.~B.}\ \bibnamefont {Garnsworthy}}, \bibinfo {author} {\bibfnamefont
  {G.~F.}\ \bibnamefont {Grinyer}}, \bibinfo {author} {\bibfnamefont
  {G.}~\bibnamefont {Hackman}}, \bibinfo {author} {\bibfnamefont
  {B.}~\bibnamefont {Jigmeddorj}}, \bibinfo {author} {\bibfnamefont
  {J.}~\bibnamefont {Jolie}}, \bibinfo {author} {\bibfnamefont {W.~D.}\
  \bibnamefont {Kulp}}, \bibinfo {author} {\bibfnamefont {K.~G.}\ \bibnamefont
  {Leach}}, \bibinfo {author} {\bibfnamefont {A.~C.}\ \bibnamefont {Morton}},
  \bibinfo {author} {\bibfnamefont {J.~N.}\ \bibnamefont {Orce}}, \bibinfo
  {author} {\bibfnamefont {C.~J.}\ \bibnamefont {Pearson}}, \bibinfo {author}
  {\bibfnamefont {A.~A.}\ \bibnamefont {Phillips}}, \bibinfo {author}
  {\bibfnamefont {A.~J.}\ \bibnamefont {Radich}}, \bibinfo {author}
  {\bibfnamefont {E.~T.}\ \bibnamefont {Rand}}, \bibinfo {author}
  {\bibfnamefont {M.~A.}\ \bibnamefont {Schumaker}}, \bibinfo {author}
  {\bibfnamefont {C.~E.}\ \bibnamefont {Svensson}}, \bibinfo {author}
  {\bibfnamefont {C.}~\bibnamefont {Sumithrarachchi}}, \bibinfo {author}
  {\bibfnamefont {S.}~\bibnamefont {Triambak}}, \bibinfo {author}
  {\bibfnamefont {N.}~\bibnamefont {Warr}}, \bibinfo {author} {\bibfnamefont
  {J.}~\bibnamefont {Wong}}, \bibinfo {author} {\bibfnamefont {J.~L.}\
  \bibnamefont {Wood}},\ and\ \bibinfo {author} {\bibfnamefont {S.~W.}\
  \bibnamefont {Yates}},\ }\bibfield  {title} {\bibinfo {title} {Shape
  coexistence and multiparticle-multihole structures in
  $^{110,112}\mathrm{Cd}$},\ }\href
  {https://doi.org/10.1103/PhysRevC.101.044302} {\bibfield  {journal} {\bibinfo
   {journal} {Phys. Rev. C}\ }\textbf {\bibinfo {volume} {101}},\ \bibinfo
  {pages} {044302} (\bibinfo {year} {2020})}\BibitemShut {NoStop}%
\bibitem [{\citenamefont {Wang}\ \emph
  {et~al.}(2023{\natexlab{a}})\citenamefont {Wang}, \citenamefont {He},\ and\
  \citenamefont {Li}}]{Wang231}%
  \BibitemOpen
  \bibfield  {author} {\bibinfo {author} {\bibfnamefont {T.}~\bibnamefont
  {Wang}}, \bibinfo {author} {\bibfnamefont {B.~C.}\ \bibnamefont {He}},\ and\
  \bibinfo {author} {\bibfnamefont {D.~K.}\ \bibnamefont {Li}},\ }\bibfield
  {title} {\bibinfo {title} {Emerging $\gamma$-softness in
  $^{196}\mathrm{Pt}$},\ }\href@noop {} {\bibfield  {journal} {\bibinfo
  {journal} {submitted}\ } (\bibinfo {year} {2023}{\natexlab{a}})}\BibitemShut
  {NoStop}%
\bibitem [{\citenamefont {Grahn}\ \emph {et~al.}(2016)\citenamefont {Grahn},
  \citenamefont {Stolze}, \citenamefont {Joss}, \citenamefont {Page},
  \citenamefont {Say\ifmmode \breve{g}\else \u{g}\fi{}\ifmmode \imath \else~\i
  \fi{}}, \citenamefont {O'Donnell}, \citenamefont {Akmali}, \citenamefont
  {Andgren}, \citenamefont {Bianco}, \citenamefont {Cullen}, \citenamefont
  {Dewald}, \citenamefont {Greenlees}, \citenamefont {Heyde}, \citenamefont
  {Iwasaki}, \citenamefont {Jakobsson}, \citenamefont {Jones}, \citenamefont
  {Judson}, \citenamefont {Julin}, \citenamefont {Juutinen}, \citenamefont
  {Ketelhut}, \citenamefont {Leino}, \citenamefont {Lumley}, \citenamefont
  {Mason}, \citenamefont {M\"oller}, \citenamefont {Nomura}, \citenamefont
  {Nyman}, \citenamefont {Petts}, \citenamefont {Peura}, \citenamefont
  {Pietralla}, \citenamefont {Pissulla}, \citenamefont {Rahkila}, \citenamefont
  {Sapple}, \citenamefont {Sar\'en}, \citenamefont {Scholey}, \citenamefont
  {Simpson}, \citenamefont {Sorri}, \citenamefont {Stevenson}, \citenamefont
  {Uusitalo}, \citenamefont {Watkins},\ and\ \citenamefont
  {Wood}}]{PhysRevC.94.044327}%
  \BibitemOpen
  \bibfield  {author} {\bibinfo {author} {\bibfnamefont {T.}~\bibnamefont
  {Grahn}}, \bibinfo {author} {\bibfnamefont {S.}~\bibnamefont {Stolze}},
  \bibinfo {author} {\bibfnamefont {D.~T.}\ \bibnamefont {Joss}}, \bibinfo
  {author} {\bibfnamefont {R.~D.}\ \bibnamefont {Page}}, \bibinfo {author}
  {\bibfnamefont {B.}~\bibnamefont {Say\ifmmode \breve{g}\else
  \u{g}\fi{}\ifmmode \imath \else~\i \fi{}}}, \bibinfo {author} {\bibfnamefont
  {D.}~\bibnamefont {O'Donnell}}, \bibinfo {author} {\bibfnamefont
  {M.}~\bibnamefont {Akmali}}, \bibinfo {author} {\bibfnamefont
  {K.}~\bibnamefont {Andgren}}, \bibinfo {author} {\bibfnamefont
  {L.}~\bibnamefont {Bianco}}, \bibinfo {author} {\bibfnamefont {D.~M.}\
  \bibnamefont {Cullen}}, \bibinfo {author} {\bibfnamefont {A.}~\bibnamefont
  {Dewald}}, \bibinfo {author} {\bibfnamefont {P.~T.}\ \bibnamefont
  {Greenlees}}, \bibinfo {author} {\bibfnamefont {K.}~\bibnamefont {Heyde}},
  \bibinfo {author} {\bibfnamefont {H.}~\bibnamefont {Iwasaki}}, \bibinfo
  {author} {\bibfnamefont {U.}~\bibnamefont {Jakobsson}}, \bibinfo {author}
  {\bibfnamefont {P.}~\bibnamefont {Jones}}, \bibinfo {author} {\bibfnamefont
  {D.~S.}\ \bibnamefont {Judson}}, \bibinfo {author} {\bibfnamefont
  {R.}~\bibnamefont {Julin}}, \bibinfo {author} {\bibfnamefont
  {S.}~\bibnamefont {Juutinen}}, \bibinfo {author} {\bibfnamefont
  {S.}~\bibnamefont {Ketelhut}}, \bibinfo {author} {\bibfnamefont
  {M.}~\bibnamefont {Leino}}, \bibinfo {author} {\bibfnamefont
  {N.}~\bibnamefont {Lumley}}, \bibinfo {author} {\bibfnamefont {P.~J.~R.}\
  \bibnamefont {Mason}}, \bibinfo {author} {\bibfnamefont {O.}~\bibnamefont
  {M\"oller}}, \bibinfo {author} {\bibfnamefont {K.}~\bibnamefont {Nomura}},
  \bibinfo {author} {\bibfnamefont {M.}~\bibnamefont {Nyman}}, \bibinfo
  {author} {\bibfnamefont {A.}~\bibnamefont {Petts}}, \bibinfo {author}
  {\bibfnamefont {P.}~\bibnamefont {Peura}}, \bibinfo {author} {\bibfnamefont
  {N.}~\bibnamefont {Pietralla}}, \bibinfo {author} {\bibfnamefont
  {T.}~\bibnamefont {Pissulla}}, \bibinfo {author} {\bibfnamefont
  {P.}~\bibnamefont {Rahkila}}, \bibinfo {author} {\bibfnamefont {P.~J.}\
  \bibnamefont {Sapple}}, \bibinfo {author} {\bibfnamefont {J.}~\bibnamefont
  {Sar\'en}}, \bibinfo {author} {\bibfnamefont {C.}~\bibnamefont {Scholey}},
  \bibinfo {author} {\bibfnamefont {J.}~\bibnamefont {Simpson}}, \bibinfo
  {author} {\bibfnamefont {J.}~\bibnamefont {Sorri}}, \bibinfo {author}
  {\bibfnamefont {P.~D.}\ \bibnamefont {Stevenson}}, \bibinfo {author}
  {\bibfnamefont {J.}~\bibnamefont {Uusitalo}}, \bibinfo {author}
  {\bibfnamefont {H.~V.}\ \bibnamefont {Watkins}},\ and\ \bibinfo {author}
  {\bibfnamefont {J.~L.}\ \bibnamefont {Wood}},\ }\bibfield  {title} {\bibinfo
  {title} {Excited states and reduced transition probabilities in
  $^{168}\mathbf{Os}$},\ }\href {https://doi.org/10.1103/PhysRevC.94.044327}
  {\bibfield  {journal} {\bibinfo  {journal} {Phys. Rev. C}\ }\textbf {\bibinfo
  {volume} {94}},\ \bibinfo {pages} {044327} (\bibinfo {year}
  {2016})}\BibitemShut {NoStop}%
\bibitem [{\citenamefont {Say\ifmmode \breve{g}\else \u{g}\fi{}\ifmmode \imath
  \else~\i \fi{}}\ \emph {et~al.}(2017)\citenamefont {Say\ifmmode
  \breve{g}\else \u{g}\fi{}\ifmmode \imath \else~\i \fi{}}, \citenamefont
  {Joss}, \citenamefont {Page}, \citenamefont {Grahn}, \citenamefont {Simpson},
  \citenamefont {O'Donnell}, \citenamefont {Alharshan}, \citenamefont
  {Auranen}, \citenamefont {B\"ack}, \citenamefont {Boening}, \citenamefont
  {Braunroth}, \citenamefont {Carroll}, \citenamefont {Cederwall},
  \citenamefont {Cullen}, \citenamefont {Dewald}, \citenamefont {Doncel},
  \citenamefont {Donosa}, \citenamefont {Drummond}, \citenamefont
  {Ertu\ifmmode~\breve{g}\else \u{g}\fi{}ral}, \citenamefont {Ert\"urk},
  \citenamefont {Fransen}, \citenamefont {Greenlees}, \citenamefont
  {Hackstein}, \citenamefont {Hauschild}, \citenamefont {Herzan}, \citenamefont
  {Jakobsson}, \citenamefont {Jones}, \citenamefont {Julin}, \citenamefont
  {Juutinen}, \citenamefont {Konki}, \citenamefont {Kr\"oll}, \citenamefont
  {Labiche}, \citenamefont {Lopez-Martens}, \citenamefont {McPeake},
  \citenamefont {Moradi}, \citenamefont {M\"oller}, \citenamefont {Mustafa},
  \citenamefont {Nieminen}, \citenamefont {Pakarinen}, \citenamefont
  {Partanen}, \citenamefont {Peura}, \citenamefont {Procter}, \citenamefont
  {Rahkila}, \citenamefont {Rother}, \citenamefont {Ruotsalainen},
  \citenamefont {Sandzelius}, \citenamefont {Sar\'en}, \citenamefont {Scholey},
  \citenamefont {Sorri}, \citenamefont {Stolze}, \citenamefont {Taylor},
  \citenamefont {Thornthwaite},\ and\ \citenamefont
  {Uusitalo}}]{PhysRevC.96.021301}%
  \BibitemOpen
  \bibfield  {author} {\bibinfo {author} {\bibfnamefont {B.}~\bibnamefont
  {Say\ifmmode \breve{g}\else \u{g}\fi{}\ifmmode \imath \else~\i \fi{}}},
  \bibinfo {author} {\bibfnamefont {D.~T.}\ \bibnamefont {Joss}}, \bibinfo
  {author} {\bibfnamefont {R.~D.}\ \bibnamefont {Page}}, \bibinfo {author}
  {\bibfnamefont {T.}~\bibnamefont {Grahn}}, \bibinfo {author} {\bibfnamefont
  {J.}~\bibnamefont {Simpson}}, \bibinfo {author} {\bibfnamefont
  {D.}~\bibnamefont {O'Donnell}}, \bibinfo {author} {\bibfnamefont
  {G.}~\bibnamefont {Alharshan}}, \bibinfo {author} {\bibfnamefont
  {K.}~\bibnamefont {Auranen}}, \bibinfo {author} {\bibfnamefont
  {T.}~\bibnamefont {B\"ack}}, \bibinfo {author} {\bibfnamefont
  {S.}~\bibnamefont {Boening}}, \bibinfo {author} {\bibfnamefont
  {T.}~\bibnamefont {Braunroth}}, \bibinfo {author} {\bibfnamefont {R.~J.}\
  \bibnamefont {Carroll}}, \bibinfo {author} {\bibfnamefont {B.}~\bibnamefont
  {Cederwall}}, \bibinfo {author} {\bibfnamefont {D.~M.}\ \bibnamefont
  {Cullen}}, \bibinfo {author} {\bibfnamefont {A.}~\bibnamefont {Dewald}},
  \bibinfo {author} {\bibfnamefont {M.}~\bibnamefont {Doncel}}, \bibinfo
  {author} {\bibfnamefont {L.}~\bibnamefont {Donosa}}, \bibinfo {author}
  {\bibfnamefont {M.~C.}\ \bibnamefont {Drummond}}, \bibinfo {author}
  {\bibfnamefont {F.}~\bibnamefont {Ertu\ifmmode~\breve{g}\else
  \u{g}\fi{}ral}}, \bibinfo {author} {\bibfnamefont {S.}~\bibnamefont
  {Ert\"urk}}, \bibinfo {author} {\bibfnamefont {C.}~\bibnamefont {Fransen}},
  \bibinfo {author} {\bibfnamefont {P.~T.}\ \bibnamefont {Greenlees}}, \bibinfo
  {author} {\bibfnamefont {M.}~\bibnamefont {Hackstein}}, \bibinfo {author}
  {\bibfnamefont {K.}~\bibnamefont {Hauschild}}, \bibinfo {author}
  {\bibfnamefont {A.}~\bibnamefont {Herzan}}, \bibinfo {author} {\bibfnamefont
  {U.}~\bibnamefont {Jakobsson}}, \bibinfo {author} {\bibfnamefont {P.~M.}\
  \bibnamefont {Jones}}, \bibinfo {author} {\bibfnamefont {R.}~\bibnamefont
  {Julin}}, \bibinfo {author} {\bibfnamefont {S.}~\bibnamefont {Juutinen}},
  \bibinfo {author} {\bibfnamefont {J.}~\bibnamefont {Konki}}, \bibinfo
  {author} {\bibfnamefont {T.}~\bibnamefont {Kr\"oll}}, \bibinfo {author}
  {\bibfnamefont {M.}~\bibnamefont {Labiche}}, \bibinfo {author} {\bibfnamefont
  {A.}~\bibnamefont {Lopez-Martens}}, \bibinfo {author} {\bibfnamefont {C.~G.}\
  \bibnamefont {McPeake}}, \bibinfo {author} {\bibfnamefont {F.}~\bibnamefont
  {Moradi}}, \bibinfo {author} {\bibfnamefont {O.}~\bibnamefont {M\"oller}},
  \bibinfo {author} {\bibfnamefont {M.}~\bibnamefont {Mustafa}}, \bibinfo
  {author} {\bibfnamefont {P.}~\bibnamefont {Nieminen}}, \bibinfo {author}
  {\bibfnamefont {J.}~\bibnamefont {Pakarinen}}, \bibinfo {author}
  {\bibfnamefont {J.}~\bibnamefont {Partanen}}, \bibinfo {author}
  {\bibfnamefont {P.}~\bibnamefont {Peura}}, \bibinfo {author} {\bibfnamefont
  {M.}~\bibnamefont {Procter}}, \bibinfo {author} {\bibfnamefont
  {P.}~\bibnamefont {Rahkila}}, \bibinfo {author} {\bibfnamefont
  {W.}~\bibnamefont {Rother}}, \bibinfo {author} {\bibfnamefont
  {P.}~\bibnamefont {Ruotsalainen}}, \bibinfo {author} {\bibfnamefont
  {M.}~\bibnamefont {Sandzelius}}, \bibinfo {author} {\bibfnamefont
  {J.}~\bibnamefont {Sar\'en}}, \bibinfo {author} {\bibfnamefont
  {C.}~\bibnamefont {Scholey}}, \bibinfo {author} {\bibfnamefont
  {J.}~\bibnamefont {Sorri}}, \bibinfo {author} {\bibfnamefont
  {S.}~\bibnamefont {Stolze}}, \bibinfo {author} {\bibfnamefont {M.~J.}\
  \bibnamefont {Taylor}}, \bibinfo {author} {\bibfnamefont {A.}~\bibnamefont
  {Thornthwaite}},\ and\ \bibinfo {author} {\bibfnamefont {J.}~\bibnamefont
  {Uusitalo}},\ }\bibfield  {title} {\bibinfo {title} {Reduced transition
  probabilities along the yrast line in $^{166}\mathrm{W}$},\ }\href
  {https://doi.org/10.1103/PhysRevC.96.021301} {\bibfield  {journal} {\bibinfo
  {journal} {Phys. Rev. C}\ }\textbf {\bibinfo {volume} {96}},\ \bibinfo
  {pages} {021301} (\bibinfo {year} {2017})}\BibitemShut {NoStop}%
\bibitem [{\citenamefont {Cederwall}\ \emph {et~al.}(2018)\citenamefont
  {Cederwall}, \citenamefont {Doncel}, \citenamefont {Aktas}, \citenamefont
  {Ertoprak}, \citenamefont {Liotta}, \citenamefont {Qi}, \citenamefont
  {Grahn}, \citenamefont {Cullen}, \citenamefont {Nara~Singh}, \citenamefont
  {Hodge}, \citenamefont {Giles}, \citenamefont {Stolze}, \citenamefont
  {Badran}, \citenamefont {Braunroth}, \citenamefont {Calverley}, \citenamefont
  {Cox}, \citenamefont {Fang}, \citenamefont {Greenlees}, \citenamefont
  {Hilton}, \citenamefont {Ideguchi}, \citenamefont {Julin}, \citenamefont
  {Juutinen}, \citenamefont {Raju}, \citenamefont {Li}, \citenamefont {Liu},
  \citenamefont {Matta}, \citenamefont {Modamio}, \citenamefont {Pakarinen},
  \citenamefont {Papadakis}, \citenamefont {Partanen}, \citenamefont
  {Petrache}, \citenamefont {Rahkila}, \citenamefont {Ruotsalainen},
  \citenamefont {Sandzelius}, \citenamefont {Sar\'en}, \citenamefont {Scholey},
  \citenamefont {Sorri}, \citenamefont {Subramaniam}, \citenamefont {Taylor},
  \citenamefont {Uusitalo},\ and\ \citenamefont
  {Valiente-Dob\'on}}]{PhysRevLett.121.022502}%
  \BibitemOpen
  \bibfield  {author} {\bibinfo {author} {\bibfnamefont {B.}~\bibnamefont
  {Cederwall}}, \bibinfo {author} {\bibfnamefont {M.}~\bibnamefont {Doncel}},
  \bibinfo {author} {\bibfnamefont {O.}~\bibnamefont {Aktas}}, \bibinfo
  {author} {\bibfnamefont {A.}~\bibnamefont {Ertoprak}}, \bibinfo {author}
  {\bibfnamefont {R.}~\bibnamefont {Liotta}}, \bibinfo {author} {\bibfnamefont
  {C.}~\bibnamefont {Qi}}, \bibinfo {author} {\bibfnamefont {T.}~\bibnamefont
  {Grahn}}, \bibinfo {author} {\bibfnamefont {D.~M.}\ \bibnamefont {Cullen}},
  \bibinfo {author} {\bibfnamefont {B.~S.}\ \bibnamefont {Nara~Singh}},
  \bibinfo {author} {\bibfnamefont {D.}~\bibnamefont {Hodge}}, \bibinfo
  {author} {\bibfnamefont {M.}~\bibnamefont {Giles}}, \bibinfo {author}
  {\bibfnamefont {S.}~\bibnamefont {Stolze}}, \bibinfo {author} {\bibfnamefont
  {H.}~\bibnamefont {Badran}}, \bibinfo {author} {\bibfnamefont
  {T.}~\bibnamefont {Braunroth}}, \bibinfo {author} {\bibfnamefont
  {T.}~\bibnamefont {Calverley}}, \bibinfo {author} {\bibfnamefont {D.~M.}\
  \bibnamefont {Cox}}, \bibinfo {author} {\bibfnamefont {Y.~D.}\ \bibnamefont
  {Fang}}, \bibinfo {author} {\bibfnamefont {P.~T.}\ \bibnamefont {Greenlees}},
  \bibinfo {author} {\bibfnamefont {J.}~\bibnamefont {Hilton}}, \bibinfo
  {author} {\bibfnamefont {E.}~\bibnamefont {Ideguchi}}, \bibinfo {author}
  {\bibfnamefont {R.}~\bibnamefont {Julin}}, \bibinfo {author} {\bibfnamefont
  {S.}~\bibnamefont {Juutinen}}, \bibinfo {author} {\bibfnamefont {M.~K.}\
  \bibnamefont {Raju}}, \bibinfo {author} {\bibfnamefont {H.}~\bibnamefont
  {Li}}, \bibinfo {author} {\bibfnamefont {H.}~\bibnamefont {Liu}}, \bibinfo
  {author} {\bibfnamefont {S.}~\bibnamefont {Matta}}, \bibinfo {author}
  {\bibfnamefont {V.}~\bibnamefont {Modamio}}, \bibinfo {author} {\bibfnamefont
  {J.}~\bibnamefont {Pakarinen}}, \bibinfo {author} {\bibfnamefont
  {P.}~\bibnamefont {Papadakis}}, \bibinfo {author} {\bibfnamefont
  {J.}~\bibnamefont {Partanen}}, \bibinfo {author} {\bibfnamefont {C.~M.}\
  \bibnamefont {Petrache}}, \bibinfo {author} {\bibfnamefont {P.}~\bibnamefont
  {Rahkila}}, \bibinfo {author} {\bibfnamefont {P.}~\bibnamefont
  {Ruotsalainen}}, \bibinfo {author} {\bibfnamefont {M.}~\bibnamefont
  {Sandzelius}}, \bibinfo {author} {\bibfnamefont {J.}~\bibnamefont {Sar\'en}},
  \bibinfo {author} {\bibfnamefont {C.}~\bibnamefont {Scholey}}, \bibinfo
  {author} {\bibfnamefont {J.}~\bibnamefont {Sorri}}, \bibinfo {author}
  {\bibfnamefont {P.}~\bibnamefont {Subramaniam}}, \bibinfo {author}
  {\bibfnamefont {M.~J.}\ \bibnamefont {Taylor}}, \bibinfo {author}
  {\bibfnamefont {J.}~\bibnamefont {Uusitalo}},\ and\ \bibinfo {author}
  {\bibfnamefont {J.~J.}\ \bibnamefont {Valiente-Dob\'on}},\ }\bibfield
  {title} {\bibinfo {title} {Lifetime measurements of excited states in
  $^{172}\mathrm{Pt}$ and the variation of quadrupole transition strength with
  angular momentum},\ }\href {https://doi.org/10.1103/PhysRevLett.121.022502}
  {\bibfield  {journal} {\bibinfo  {journal} {Phys. Rev. Lett.}\ }\textbf
  {\bibinfo {volume} {121}},\ \bibinfo {pages} {022502} (\bibinfo {year}
  {2018})}\BibitemShut {NoStop}%
\bibitem [{\citenamefont {Goasduff}\ \emph {et~al.}(2019)\citenamefont
  {Goasduff}, \citenamefont {Ljungvall}, \citenamefont {Rodr\'{\i}guez},
  \citenamefont {Bello~Garrote}, \citenamefont {Etile}, \citenamefont
  {Georgiev}, \citenamefont {Giacoppo}, \citenamefont {Grente}, \citenamefont
  {Klintefjord}, \citenamefont {Ku\ifmmode \mbox{\c{s}}\else
  \c{s}\fi{}o\ifmmode~\breve{g}\else \u{g}\fi{}lu}, \citenamefont {Matea},
  \citenamefont {Roccia}, \citenamefont {Salsac},\ and\ \citenamefont
  {Sotty}}]{PhysRevC.100.034302}%
  \BibitemOpen
  \bibfield  {author} {\bibinfo {author} {\bibfnamefont {A.}~\bibnamefont
  {Goasduff}}, \bibinfo {author} {\bibfnamefont {J.}~\bibnamefont {Ljungvall}},
  \bibinfo {author} {\bibfnamefont {T.~R.}\ \bibnamefont {Rodr\'{\i}guez}},
  \bibinfo {author} {\bibfnamefont {F.~L.}\ \bibnamefont {Bello~Garrote}},
  \bibinfo {author} {\bibfnamefont {A.}~\bibnamefont {Etile}}, \bibinfo
  {author} {\bibfnamefont {G.}~\bibnamefont {Georgiev}}, \bibinfo {author}
  {\bibfnamefont {F.}~\bibnamefont {Giacoppo}}, \bibinfo {author}
  {\bibfnamefont {L.}~\bibnamefont {Grente}}, \bibinfo {author} {\bibfnamefont
  {M.}~\bibnamefont {Klintefjord}}, \bibinfo {author} {\bibfnamefont
  {A.}~\bibnamefont {Ku\ifmmode \mbox{\c{s}}\else
  \c{s}\fi{}o\ifmmode~\breve{g}\else \u{g}\fi{}lu}}, \bibinfo {author}
  {\bibfnamefont {I.}~\bibnamefont {Matea}}, \bibinfo {author} {\bibfnamefont
  {S.}~\bibnamefont {Roccia}}, \bibinfo {author} {\bibfnamefont {M.-D.}\
  \bibnamefont {Salsac}},\ and\ \bibinfo {author} {\bibfnamefont
  {C.}~\bibnamefont {Sotty}},\ }\bibfield  {title} {\bibinfo {title} {{B(E2)}
  anomalies in the yrast band of $^{170}\mathrm{Os}$},\ }\href
  {https://doi.org/10.1103/PhysRevC.100.034302} {\bibfield  {journal} {\bibinfo
   {journal} {Phys. Rev. C}\ }\textbf {\bibinfo {volume} {100}},\ \bibinfo
  {pages} {034302} (\bibinfo {year} {2019})}\BibitemShut {NoStop}%
\bibitem [{\citenamefont {Wang}(2020)}]{Wang2020}%
  \BibitemOpen
  \bibfield  {author} {\bibinfo {author} {\bibfnamefont {T.}~\bibnamefont
  {Wang}},\ }\bibfield  {title} {\bibinfo {title} {A collective description of
  the unusually low ratio ${B}_{4/2}$ = {B(E2;$4_{1}^{+}\rightarrow
  2_{1}^{+}$)/B(E2;$2_{1}^{+}\rightarrow 0_{1}^{+})$}},\ }\href
  {https://doi.org/10.1209/0295-5075/129/52001} {\bibfield  {journal} {\bibinfo
   {journal} {EPL}\ }\textbf {\bibinfo {volume} {129}},\ \bibinfo {pages}
  {52001} (\bibinfo {year} {2020})}\BibitemShut {NoStop}%
\bibitem [{\citenamefont {Zhang}\ \emph
  {et~al.}(2022{\natexlab{a}})\citenamefont {Zhang}, \citenamefont {He},
  \citenamefont {Karlsson}, \citenamefont {Qi}, \citenamefont {Pan},\ and\
  \citenamefont {Draayer}}]{Yu2022}%
  \BibitemOpen
  \bibfield  {author} {\bibinfo {author} {\bibfnamefont {Y.}~\bibnamefont
  {Zhang}}, \bibinfo {author} {\bibfnamefont {Y.-W.}\ \bibnamefont {He}},
  \bibinfo {author} {\bibfnamefont {D.}~\bibnamefont {Karlsson}}, \bibinfo
  {author} {\bibfnamefont {C.}~\bibnamefont {Qi}}, \bibinfo {author}
  {\bibfnamefont {F.}~\bibnamefont {Pan}},\ and\ \bibinfo {author}
  {\bibfnamefont {J.}~\bibnamefont {Draayer}},\ }\bibfield  {title} {\bibinfo
  {title} {A theoretical interpretation of the anomalous reduced {E}2
  transition probabilities along the yrast line of neutron-deficient nuclei},\
  }\href {https://doi.org/https://doi.org/10.1016/j.physletb.2022.137443}
  {\bibfield  {journal} {\bibinfo  {journal} {Phys. Lett. B}\ }\textbf
  {\bibinfo {volume} {834}},\ \bibinfo {pages} {137443} (\bibinfo {year}
  {2022}{\natexlab{a}})}\BibitemShut {NoStop}%
\bibitem [{\citenamefont {Wang}\ \emph
  {et~al.}(2023{\natexlab{b}})\citenamefont {Wang}, \citenamefont {He},\ and\
  \citenamefont {Li}}]{Wang232}%
  \BibitemOpen
  \bibfield  {author} {\bibinfo {author} {\bibfnamefont {T.}~\bibnamefont
  {Wang}}, \bibinfo {author} {\bibfnamefont {B.~C.}\ \bibnamefont {He}},\ and\
  \bibinfo {author} {\bibfnamefont {D.~K.}\ \bibnamefont {Li}},\ }\bibfield
  {title} {\bibinfo {title} {Prolate-oblate asymmetric shape phase transition
  in the interacting boson model with {SU}(3) higher-order interaction},\
  }\href@noop {} {\bibfield  {journal} {\bibinfo  {journal} {submitted}\ }
  (\bibinfo {year} {2023}{\natexlab{b}})}\BibitemShut {NoStop}%
\bibitem [{\citenamefont {Kaneko}\ \emph {et~al.}(2023)\citenamefont {Kaneko},
  \citenamefont {Sun}, \citenamefont {Shimizu},\ and\ \citenamefont
  {Mizusaki}}]{PhysRevLett.130.052501}%
  \BibitemOpen
  \bibfield  {author} {\bibinfo {author} {\bibfnamefont {K.}~\bibnamefont
  {Kaneko}}, \bibinfo {author} {\bibfnamefont {Y.}~\bibnamefont {Sun}},
  \bibinfo {author} {\bibfnamefont {N.}~\bibnamefont {Shimizu}},\ and\ \bibinfo
  {author} {\bibfnamefont {T.}~\bibnamefont {Mizusaki}},\ }\bibfield  {title}
  {\bibinfo {title} {Quasi-{SU}(3) coupling induced oblate-prolate shape phase
  transition in the casten triangle},\ }\href
  {https://doi.org/10.1103/PhysRevLett.130.052501} {\bibfield  {journal}
  {\bibinfo  {journal} {Phys. Rev. Lett.}\ }\textbf {\bibinfo {volume} {130}},\
  \bibinfo {pages} {052501} (\bibinfo {year} {2023})}\BibitemShut {NoStop}%
\bibitem [{\citenamefont {Rajbanshi}\ \emph {et~al.}(2021)\citenamefont
  {Rajbanshi}, \citenamefont {Bhattacharya}, \citenamefont {Raut},
  \citenamefont {Palit}, \citenamefont {Ali}, \citenamefont {Santra},
  \citenamefont {Pai}, \citenamefont {Babra}, \citenamefont {Banik},
  \citenamefont {Bhattacharyya}, \citenamefont {Dey}, \citenamefont
  {Mukherjee}, \citenamefont {Laskar}, \citenamefont {Nandi}, \citenamefont
  {Trivedi}, \citenamefont {Ghugre},\ and\ \citenamefont
  {Goswami}}]{Rajban2021}%
  \BibitemOpen
  \bibfield  {author} {\bibinfo {author} {\bibfnamefont {S.}~\bibnamefont
  {Rajbanshi}}, \bibinfo {author} {\bibfnamefont {S.}~\bibnamefont
  {Bhattacharya}}, \bibinfo {author} {\bibfnamefont {R.}~\bibnamefont {Raut}},
  \bibinfo {author} {\bibfnamefont {R.}~\bibnamefont {Palit}}, \bibinfo
  {author} {\bibfnamefont {S.}~\bibnamefont {Ali}}, \bibinfo {author}
  {\bibfnamefont {R.}~\bibnamefont {Santra}}, \bibinfo {author} {\bibfnamefont
  {H.}~\bibnamefont {Pai}}, \bibinfo {author} {\bibfnamefont {F.~S.}\
  \bibnamefont {Babra}}, \bibinfo {author} {\bibfnamefont {R.}~\bibnamefont
  {Banik}}, \bibinfo {author} {\bibfnamefont {S.}~\bibnamefont
  {Bhattacharyya}}, \bibinfo {author} {\bibfnamefont {P.}~\bibnamefont {Dey}},
  \bibinfo {author} {\bibfnamefont {G.}~\bibnamefont {Mukherjee}}, \bibinfo
  {author} {\bibfnamefont {M.~S.~R.}\ \bibnamefont {Laskar}}, \bibinfo {author}
  {\bibfnamefont {S.}~\bibnamefont {Nandi}}, \bibinfo {author} {\bibfnamefont
  {T.}~\bibnamefont {Trivedi}}, \bibinfo {author} {\bibfnamefont {S.~S.}\
  \bibnamefont {Ghugre}},\ and\ \bibinfo {author} {\bibfnamefont
  {A.}~\bibnamefont {Goswami}},\ }\bibfield  {title} {\bibinfo {title}
  {Experimental evidence of exact {E}(5) symmetry in $^{82}\mathrm{Kr}$},\
  }\href {https://doi.org/10.1103/PhysRevC.104.L031302} {\bibfield  {journal}
  {\bibinfo  {journal} {Phys. Rev. C}\ }\textbf {\bibinfo {volume} {104}},\
  \bibinfo {pages} {L031302} (\bibinfo {year} {2021})}\BibitemShut {NoStop}%
\bibitem [{\citenamefont {L\'opez-Moreno}\ and\ \citenamefont
  {Casta\~nos}(1996)}]{PhysRevC.54.2374}%
  \BibitemOpen
  \bibfield  {author} {\bibinfo {author} {\bibfnamefont {E.}~\bibnamefont
  {L\'opez-Moreno}}\ and\ \bibinfo {author} {\bibfnamefont {O.}~\bibnamefont
  {Casta\~nos}},\ }\bibfield  {title} {\bibinfo {title} {Shapes and stability
  within the interacting boson model: Dynamical symmetries},\ }\href
  {https://doi.org/10.1103/PhysRevC.54.2374} {\bibfield  {journal} {\bibinfo
  {journal} {Phys. Rev. C}\ }\textbf {\bibinfo {volume} {54}},\ \bibinfo
  {pages} {2374} (\bibinfo {year} {1996})}\BibitemShut {NoStop}%
\bibitem [{\citenamefont {Jolie}\ \emph {et~al.}(2001)\citenamefont {Jolie},
  \citenamefont {Casten}, \citenamefont {von Brentano},\ and\ \citenamefont
  {Werner}}]{PhysRevLett.87.162501}%
  \BibitemOpen
  \bibfield  {author} {\bibinfo {author} {\bibfnamefont {J.}~\bibnamefont
  {Jolie}}, \bibinfo {author} {\bibfnamefont {R.~F.}\ \bibnamefont {Casten}},
  \bibinfo {author} {\bibfnamefont {P.}~\bibnamefont {von Brentano}},\ and\
  \bibinfo {author} {\bibfnamefont {V.}~\bibnamefont {Werner}},\ }\bibfield
  {title} {\bibinfo {title} {Quantum phase transition for
  $\mathit{\ensuremath{\gamma}}$-soft nuclei},\ }\href
  {https://doi.org/10.1103/PhysRevLett.87.162501} {\bibfield  {journal}
  {\bibinfo  {journal} {Phys. Rev. Lett.}\ }\textbf {\bibinfo {volume} {87}},\
  \bibinfo {pages} {162501} (\bibinfo {year} {2001})}\BibitemShut {NoStop}%
\bibitem [{\citenamefont {Jolie}\ and\ \citenamefont
  {Linnemann}(2003)}]{PhysRevC.68.031301}%
  \BibitemOpen
  \bibfield  {author} {\bibinfo {author} {\bibfnamefont {J.}~\bibnamefont
  {Jolie}}\ and\ \bibinfo {author} {\bibfnamefont {A.}~\bibnamefont
  {Linnemann}},\ }\bibfield  {title} {\bibinfo {title} {Prolate-oblate phase
  transition in the $\mathrm{Hf-Hg}$ mass region},\ }\href
  {https://doi.org/10.1103/PhysRevC.68.031301} {\bibfield  {journal} {\bibinfo
  {journal} {Phys. Rev. C}\ }\textbf {\bibinfo {volume} {68}},\ \bibinfo
  {pages} {031301} (\bibinfo {year} {2003})}\BibitemShut {NoStop}%
\bibitem [{\citenamefont {Karakatsanis}\ and\ \citenamefont
  {Nomura}(2022)}]{Nomura2022}%
  \BibitemOpen
  \bibfield  {author} {\bibinfo {author} {\bibfnamefont {K.~E.}\ \bibnamefont
  {Karakatsanis}}\ and\ \bibinfo {author} {\bibfnamefont {K.}~\bibnamefont
  {Nomura}},\ }\bibfield  {title} {\bibinfo {title} {Signatures of shape phase
  transitions in krypton isotopes based on relativistic energy density
  functionals},\ }\href {https://doi.org/10.1103/PhysRevC.105.064310}
  {\bibfield  {journal} {\bibinfo  {journal} {Phys. Rev. C}\ }\textbf {\bibinfo
  {volume} {105}},\ \bibinfo {pages} {064310} (\bibinfo {year}
  {2022})}\BibitemShut {NoStop}%
\bibitem [{\citenamefont {Rowe}(1985)}]{Rowe1985}%
  \BibitemOpen
  \bibfield  {author} {\bibinfo {author} {\bibfnamefont {D.~J.}\ \bibnamefont
  {Rowe}},\ }\bibfield  {title} {\bibinfo {title} {Microscopic theory of the
  nuclear collective model},\ }\href
  {https://doi.org/10.1088/0034-4885/48/10/003} {\bibfield  {journal} {\bibinfo
   {journal} {Rep. Prog. Phys.}\ }\textbf {\bibinfo {volume} {48}},\ \bibinfo
  {pages} {1419} (\bibinfo {year} {1985})}\BibitemShut {NoStop}%
\bibitem [{\citenamefont {Zhang}\ \emph
  {et~al.}(2022{\natexlab{b}})\citenamefont {Zhang}, \citenamefont {Wang},
  \citenamefont {Jiang},\ and\ \citenamefont {Chen}}]{sym14102219}%
  \BibitemOpen
  \bibfield  {author} {\bibinfo {author} {\bibfnamefont {Y.}~\bibnamefont
  {Zhang}}, \bibinfo {author} {\bibfnamefont {Z.-T.}\ \bibnamefont {Wang}},
  \bibinfo {author} {\bibfnamefont {H.-D.}\ \bibnamefont {Jiang}},\ and\
  \bibinfo {author} {\bibfnamefont {X.}~\bibnamefont {Chen}},\ }\bibfield
  {title} {\bibinfo {title} {Hidden euclidean dynamical symmetry in the {U}(n +
  1) vibron model},\ }\bibfield  {journal} {\bibinfo  {journal} {Symmetry}\
  }\textbf {\bibinfo {volume} {14}},\ \href
  {https://doi.org/10.3390/sym14102219} {10.3390/sym14102219} (\bibinfo {year}
  {2022}{\natexlab{b}})\BibitemShut {NoStop}%
\bibitem [{\citenamefont {Coquard}\ \emph {et~al.}(2009)\citenamefont
  {Coquard}, \citenamefont {Pietralla}, \citenamefont {Ahn}, \citenamefont
  {Rainovski}, \citenamefont {Bettermann}, \citenamefont {Carpenter},
  \citenamefont {Janssens}, \citenamefont {Leske}, \citenamefont {Lister},
  \citenamefont {M\"oller}, \citenamefont {Rother}, \citenamefont {Werner},\
  and\ \citenamefont {Zhu}}]{PhysRevC.80.061304}%
  \BibitemOpen
  \bibfield  {author} {\bibinfo {author} {\bibfnamefont {L.}~\bibnamefont
  {Coquard}}, \bibinfo {author} {\bibfnamefont {N.}~\bibnamefont {Pietralla}},
  \bibinfo {author} {\bibfnamefont {T.}~\bibnamefont {Ahn}}, \bibinfo {author}
  {\bibfnamefont {G.}~\bibnamefont {Rainovski}}, \bibinfo {author}
  {\bibfnamefont {L.}~\bibnamefont {Bettermann}}, \bibinfo {author}
  {\bibfnamefont {M.~P.}\ \bibnamefont {Carpenter}}, \bibinfo {author}
  {\bibfnamefont {R.~V.~F.}\ \bibnamefont {Janssens}}, \bibinfo {author}
  {\bibfnamefont {J.}~\bibnamefont {Leske}}, \bibinfo {author} {\bibfnamefont
  {C.~J.}\ \bibnamefont {Lister}}, \bibinfo {author} {\bibfnamefont
  {O.}~\bibnamefont {M\"oller}}, \bibinfo {author} {\bibfnamefont
  {W.}~\bibnamefont {Rother}}, \bibinfo {author} {\bibfnamefont
  {V.}~\bibnamefont {Werner}},\ and\ \bibinfo {author} {\bibfnamefont
  {S.}~\bibnamefont {Zhu}},\ }\bibfield  {title} {\bibinfo {title} {Robust test
  of {E}(5) symmetry in $^{128}\mathrm{Xe}$},\ }\href
  {https://doi.org/10.1103/PhysRevC.80.061304} {\bibfield  {journal} {\bibinfo
  {journal} {Phys. Rev. C}\ }\textbf {\bibinfo {volume} {80}},\ \bibinfo
  {pages} {061304} (\bibinfo {year} {2009})}\BibitemShut {NoStop}%
\bibitem [{\citenamefont {Coquard}\ \emph {et~al.}(2011)\citenamefont
  {Coquard}, \citenamefont {Rainovski}, \citenamefont {Pietralla},
  \citenamefont {Ahn}, \citenamefont {Bettermann}, \citenamefont {Carpenter},
  \citenamefont {Janssens}, \citenamefont {Leske}, \citenamefont {Lister},
  \citenamefont {M\"oller}, \citenamefont {M\"oller}, \citenamefont {Rother},
  \citenamefont {Werner},\ and\ \citenamefont {Zhu}}]{PhysRevC.83.044318}%
  \BibitemOpen
  \bibfield  {author} {\bibinfo {author} {\bibfnamefont {L.}~\bibnamefont
  {Coquard}}, \bibinfo {author} {\bibfnamefont {G.}~\bibnamefont {Rainovski}},
  \bibinfo {author} {\bibfnamefont {N.}~\bibnamefont {Pietralla}}, \bibinfo
  {author} {\bibfnamefont {T.}~\bibnamefont {Ahn}}, \bibinfo {author}
  {\bibfnamefont {L.}~\bibnamefont {Bettermann}}, \bibinfo {author}
  {\bibfnamefont {M.~P.}\ \bibnamefont {Carpenter}}, \bibinfo {author}
  {\bibfnamefont {R.~V.~F.}\ \bibnamefont {Janssens}}, \bibinfo {author}
  {\bibfnamefont {J.}~\bibnamefont {Leske}}, \bibinfo {author} {\bibfnamefont
  {C.~J.}\ \bibnamefont {Lister}}, \bibinfo {author} {\bibfnamefont
  {O.}~\bibnamefont {M\"oller}}, \bibinfo {author} {\bibfnamefont
  {T.}~\bibnamefont {M\"oller}}, \bibinfo {author} {\bibfnamefont
  {W.}~\bibnamefont {Rother}}, \bibinfo {author} {\bibfnamefont
  {V.}~\bibnamefont {Werner}},\ and\ \bibinfo {author} {\bibfnamefont
  {S.}~\bibnamefont {Zhu}},\ }\bibfield  {title} {\bibinfo {title}
  {O(6)-symmetry breaking in the $\ensuremath{\gamma}$-soft nucleus
  $^{126}\mathrm{Xe}$ and its evolution in the light stable xenon isotopes},\
  }\href {https://doi.org/10.1103/PhysRevC.83.044318} {\bibfield  {journal}
  {\bibinfo  {journal} {Phys. Rev. C}\ }\textbf {\bibinfo {volume} {83}},\
  \bibinfo {pages} {044318} (\bibinfo {year} {2011})}\BibitemShut {NoStop}%
\bibitem [{\citenamefont {Peters}\ \emph {et~al.}(2016)\citenamefont {Peters},
  \citenamefont {Ross}, \citenamefont {Ashley}, \citenamefont {Chakraborty},
  \citenamefont {Crider}, \citenamefont {Hennek}, \citenamefont {Liu},
  \citenamefont {McEllistrem}, \citenamefont {Mukhopadhyay}, \citenamefont
  {Prados-Est\'evez}, \citenamefont {Ramirez}, \citenamefont {Thrasher},\ and\
  \citenamefont {Yates}}]{PhysRevC.94.024313}%
  \BibitemOpen
  \bibfield  {author} {\bibinfo {author} {\bibfnamefont {E.~E.}\ \bibnamefont
  {Peters}}, \bibinfo {author} {\bibfnamefont {T.~J.}\ \bibnamefont {Ross}},
  \bibinfo {author} {\bibfnamefont {S.~F.}\ \bibnamefont {Ashley}}, \bibinfo
  {author} {\bibfnamefont {A.}~\bibnamefont {Chakraborty}}, \bibinfo {author}
  {\bibfnamefont {B.~P.}\ \bibnamefont {Crider}}, \bibinfo {author}
  {\bibfnamefont {M.~D.}\ \bibnamefont {Hennek}}, \bibinfo {author}
  {\bibfnamefont {S.~H.}\ \bibnamefont {Liu}}, \bibinfo {author} {\bibfnamefont
  {M.~T.}\ \bibnamefont {McEllistrem}}, \bibinfo {author} {\bibfnamefont
  {S.}~\bibnamefont {Mukhopadhyay}}, \bibinfo {author} {\bibfnamefont {F.~M.}\
  \bibnamefont {Prados-Est\'evez}}, \bibinfo {author} {\bibfnamefont
  {A.~P.~D.}\ \bibnamefont {Ramirez}}, \bibinfo {author} {\bibfnamefont
  {J.~S.}\ \bibnamefont {Thrasher}},\ and\ \bibinfo {author} {\bibfnamefont
  {S.~W.}\ \bibnamefont {Yates}},\ }\bibfield  {title} {\bibinfo {title}
  {${0}^{+}$ states in $^{130,132}\mathrm{Xe}$: A search for {E}(5) behavior},\
  }\href {https://doi.org/10.1103/PhysRevC.94.024313} {\bibfield  {journal}
  {\bibinfo  {journal} {Phys. Rev. C}\ }\textbf {\bibinfo {volume} {94}},\
  \bibinfo {pages} {024313} (\bibinfo {year} {2016})}\BibitemShut {NoStop}%
\bibitem [{\citenamefont {Rainovski}\ \emph {et~al.}(2010)\citenamefont
  {Rainovski}, \citenamefont {Pietralla}, \citenamefont {Ahn}, \citenamefont
  {Coquard}, \citenamefont {Lister}, \citenamefont {Janssens}, \citenamefont
  {Carpenter}, \citenamefont {Zhu}, \citenamefont {Bettermann}, \citenamefont
  {Jolie}, \citenamefont {Rother}, \citenamefont {Jolos},\ and\ \citenamefont
  {Werner}}]{RAINOVSKI201011}%
  \BibitemOpen
  \bibfield  {author} {\bibinfo {author} {\bibfnamefont {G.}~\bibnamefont
  {Rainovski}}, \bibinfo {author} {\bibfnamefont {N.}~\bibnamefont
  {Pietralla}}, \bibinfo {author} {\bibfnamefont {T.}~\bibnamefont {Ahn}},
  \bibinfo {author} {\bibfnamefont {L.}~\bibnamefont {Coquard}}, \bibinfo
  {author} {\bibfnamefont {C.}~\bibnamefont {Lister}}, \bibinfo {author}
  {\bibfnamefont {R.}~\bibnamefont {Janssens}}, \bibinfo {author}
  {\bibfnamefont {M.}~\bibnamefont {Carpenter}}, \bibinfo {author}
  {\bibfnamefont {S.}~\bibnamefont {Zhu}}, \bibinfo {author} {\bibfnamefont
  {L.}~\bibnamefont {Bettermann}}, \bibinfo {author} {\bibfnamefont
  {J.}~\bibnamefont {Jolie}}, \bibinfo {author} {\bibfnamefont
  {W.}~\bibnamefont {Rother}}, \bibinfo {author} {\bibfnamefont
  {R.}~\bibnamefont {Jolos}},\ and\ \bibinfo {author} {\bibfnamefont
  {V.}~\bibnamefont {Werner}},\ }\bibfield  {title} {\bibinfo {title} {How
  close to the {O}(6) symmetry is the nucleus $^{124}\mathrm{Xe}$?},\ }\href
  {https://doi.org/https://doi.org/10.1016/j.physletb.2009.12.007} {\bibfield
  {journal} {\bibinfo  {journal} {Physics Letters B}\ }\textbf {\bibinfo
  {volume} {683}},\ \bibinfo {pages} {11} (\bibinfo {year} {2010})}\BibitemShut
  {NoStop}%
\bibitem [{\citenamefont {Morrison}\ \emph {et~al.}(2020)\citenamefont
  {Morrison}, \citenamefont {Hady\ifmmode \acute{n}\else
  \'{n}\fi{}ska-Kl\ifmmode~\mbox{\c{e}}\else \c{e}\fi{}k}, \citenamefont
  {Podoly\'ak}, \citenamefont {Doherty}, \citenamefont {Gaffney}, \citenamefont
  {Kaya}, \citenamefont {Pr\'ochniak}, \citenamefont
  {Samorajczyk-Py\ifmmode~\acute{s}\else \'{s}\fi{}k}, \citenamefont {Srebrny},
  \citenamefont {Berry}, \citenamefont {Boukhari}, \citenamefont {Brunet},
  \citenamefont {Canavan}, \citenamefont {Catherall}, \citenamefont {Colosimo},
  \citenamefont {Cubiss}, \citenamefont {De~Witte}, \citenamefont {Fransen},
  \citenamefont {Giannopoulos}, \citenamefont {Hess}, \citenamefont {Kr\"oll},
  \citenamefont {Lalovi\ifmmode~\acute{c}\else \'{c}\fi{}}, \citenamefont
  {Marsh}, \citenamefont {Palenzuela}, \citenamefont {Napiorkowski},
  \citenamefont {O'Neill}, \citenamefont {Pakarinen}, \citenamefont {Ramos},
  \citenamefont {Reiter}, \citenamefont {Rodriguez}, \citenamefont {Rosiak},
  \citenamefont {Rothe}, \citenamefont {Rudigier}, \citenamefont {Siciliano},
  \citenamefont {Sn\"all}, \citenamefont {Spagnoletti}, \citenamefont {Thiel},
  \citenamefont {Warr}, \citenamefont {Wenander}, \citenamefont {Zidarova},\
  and\ \citenamefont {Zieli\ifmmode~\acute{n}\else
  \'{n}\fi{}ska}}]{PhysRevC.102.054304}%
  \BibitemOpen
  \bibfield  {author} {\bibinfo {author} {\bibfnamefont {L.}~\bibnamefont
  {Morrison}}, \bibinfo {author} {\bibfnamefont {K.}~\bibnamefont {Hady\ifmmode
  \acute{n}\else \'{n}\fi{}ska-Kl\ifmmode~\mbox{\c{e}}\else \c{e}\fi{}k}},
  \bibinfo {author} {\bibfnamefont {Z.}~\bibnamefont {Podoly\'ak}}, \bibinfo
  {author} {\bibfnamefont {D.~T.}\ \bibnamefont {Doherty}}, \bibinfo {author}
  {\bibfnamefont {L.~P.}\ \bibnamefont {Gaffney}}, \bibinfo {author}
  {\bibfnamefont {L.}~\bibnamefont {Kaya}}, \bibinfo {author} {\bibfnamefont
  {L.}~\bibnamefont {Pr\'ochniak}}, \bibinfo {author} {\bibfnamefont
  {J.}~\bibnamefont {Samorajczyk-Py\ifmmode~\acute{s}\else \'{s}\fi{}k}},
  \bibinfo {author} {\bibfnamefont {J.}~\bibnamefont {Srebrny}}, \bibinfo
  {author} {\bibfnamefont {T.}~\bibnamefont {Berry}}, \bibinfo {author}
  {\bibfnamefont {A.}~\bibnamefont {Boukhari}}, \bibinfo {author}
  {\bibfnamefont {M.}~\bibnamefont {Brunet}}, \bibinfo {author} {\bibfnamefont
  {R.}~\bibnamefont {Canavan}}, \bibinfo {author} {\bibfnamefont
  {R.}~\bibnamefont {Catherall}}, \bibinfo {author} {\bibfnamefont {S.~J.}\
  \bibnamefont {Colosimo}}, \bibinfo {author} {\bibfnamefont {J.~G.}\
  \bibnamefont {Cubiss}}, \bibinfo {author} {\bibfnamefont {H.}~\bibnamefont
  {De~Witte}}, \bibinfo {author} {\bibfnamefont {C.}~\bibnamefont {Fransen}},
  \bibinfo {author} {\bibfnamefont {E.}~\bibnamefont {Giannopoulos}}, \bibinfo
  {author} {\bibfnamefont {H.}~\bibnamefont {Hess}}, \bibinfo {author}
  {\bibfnamefont {T.}~\bibnamefont {Kr\"oll}}, \bibinfo {author} {\bibfnamefont
  {N.}~\bibnamefont {Lalovi\ifmmode~\acute{c}\else \'{c}\fi{}}}, \bibinfo
  {author} {\bibfnamefont {B.}~\bibnamefont {Marsh}}, \bibinfo {author}
  {\bibfnamefont {Y.~M.}\ \bibnamefont {Palenzuela}}, \bibinfo {author}
  {\bibfnamefont {P.~J.}\ \bibnamefont {Napiorkowski}}, \bibinfo {author}
  {\bibfnamefont {G.}~\bibnamefont {O'Neill}}, \bibinfo {author} {\bibfnamefont
  {J.}~\bibnamefont {Pakarinen}}, \bibinfo {author} {\bibfnamefont {J.~P.}\
  \bibnamefont {Ramos}}, \bibinfo {author} {\bibfnamefont {P.}~\bibnamefont
  {Reiter}}, \bibinfo {author} {\bibfnamefont {J.~A.}\ \bibnamefont
  {Rodriguez}}, \bibinfo {author} {\bibfnamefont {D.}~\bibnamefont {Rosiak}},
  \bibinfo {author} {\bibfnamefont {S.}~\bibnamefont {Rothe}}, \bibinfo
  {author} {\bibfnamefont {M.}~\bibnamefont {Rudigier}}, \bibinfo {author}
  {\bibfnamefont {M.}~\bibnamefont {Siciliano}}, \bibinfo {author}
  {\bibfnamefont {J.}~\bibnamefont {Sn\"all}}, \bibinfo {author} {\bibfnamefont
  {P.}~\bibnamefont {Spagnoletti}}, \bibinfo {author} {\bibfnamefont
  {S.}~\bibnamefont {Thiel}}, \bibinfo {author} {\bibfnamefont
  {N.}~\bibnamefont {Warr}}, \bibinfo {author} {\bibfnamefont {F.}~\bibnamefont
  {Wenander}}, \bibinfo {author} {\bibfnamefont {R.}~\bibnamefont {Zidarova}},\
  and\ \bibinfo {author} {\bibfnamefont {M.}~\bibnamefont
  {Zieli\ifmmode~\acute{n}\else \'{n}\fi{}ska}},\ }\bibfield  {title} {\bibinfo
  {title} {Quadrupole deformation of $^{130}\mathrm{Xe}$ measured in a
  {C}oulomb-excitation experiment},\ }\href
  {https://doi.org/10.1103/PhysRevC.102.054304} {\bibfield  {journal} {\bibinfo
   {journal} {Phys. Rev. C}\ }\textbf {\bibinfo {volume} {102}},\ \bibinfo
  {pages} {054304} (\bibinfo {year} {2020})}\BibitemShut {NoStop}%
\bibitem [{\citenamefont {Kisyov}\ \emph {et~al.}(2022)\citenamefont {Kisyov},
  \citenamefont {Wu}, \citenamefont {Henderson}, \citenamefont {Gade},
  \citenamefont {Kaneko}, \citenamefont {Sun}, \citenamefont {Shimizu},
  \citenamefont {Mizusaki}, \citenamefont {Rhodes}, \citenamefont {Biswas},
  \citenamefont {Chester}, \citenamefont {Devlin}, \citenamefont {Farris},
  \citenamefont {Hill}, \citenamefont {Li}, \citenamefont {Rubino},\ and\
  \citenamefont {Weisshaar}}]{PhysRevC.106.034311}%
  \BibitemOpen
  \bibfield  {author} {\bibinfo {author} {\bibfnamefont {S.}~\bibnamefont
  {Kisyov}}, \bibinfo {author} {\bibfnamefont {C.~Y.}\ \bibnamefont {Wu}},
  \bibinfo {author} {\bibfnamefont {J.}~\bibnamefont {Henderson}}, \bibinfo
  {author} {\bibfnamefont {A.}~\bibnamefont {Gade}}, \bibinfo {author}
  {\bibfnamefont {K.}~\bibnamefont {Kaneko}}, \bibinfo {author} {\bibfnamefont
  {Y.}~\bibnamefont {Sun}}, \bibinfo {author} {\bibfnamefont {N.}~\bibnamefont
  {Shimizu}}, \bibinfo {author} {\bibfnamefont {T.}~\bibnamefont {Mizusaki}},
  \bibinfo {author} {\bibfnamefont {D.}~\bibnamefont {Rhodes}}, \bibinfo
  {author} {\bibfnamefont {S.}~\bibnamefont {Biswas}}, \bibinfo {author}
  {\bibfnamefont {A.}~\bibnamefont {Chester}}, \bibinfo {author} {\bibfnamefont
  {M.}~\bibnamefont {Devlin}}, \bibinfo {author} {\bibfnamefont
  {P.}~\bibnamefont {Farris}}, \bibinfo {author} {\bibfnamefont {A.~M.}\
  \bibnamefont {Hill}}, \bibinfo {author} {\bibfnamefont {J.}~\bibnamefont
  {Li}}, \bibinfo {author} {\bibfnamefont {E.}~\bibnamefont {Rubino}},\ and\
  \bibinfo {author} {\bibfnamefont {D.}~\bibnamefont {Weisshaar}},\ }\bibfield
  {title} {\bibinfo {title} {Structure of $^{126,128}\mathrm{Xe}$ studied in
  {C}oulomb excitation measurements},\ }\href
  {https://doi.org/10.1103/PhysRevC.106.034311} {\bibfield  {journal} {\bibinfo
   {journal} {Phys. Rev. C}\ }\textbf {\bibinfo {volume} {106}},\ \bibinfo
  {pages} {034311} (\bibinfo {year} {2022})}\BibitemShut {NoStop}%
\bibitem [{\citenamefont {Cl\'ement}\ \emph {et~al.}(2023)\citenamefont
  {Cl\'ement}, \citenamefont {Lemasson}, \citenamefont {Rejmund}, \citenamefont
  {Jacquot}, \citenamefont {Ralet}, \citenamefont {Michelagnoli}, \citenamefont
  {Barrientos}, \citenamefont {Bednarczyk}, \citenamefont {Benzoni},
  \citenamefont {Boston}, \citenamefont {Bracco}, \citenamefont {Cederwall},
  \citenamefont {Ciemala}, \citenamefont {Collado}, \citenamefont {Crespi},
  \citenamefont {Domingo-Pardo}, \citenamefont {Dudouet}, \citenamefont
  {Eberth}, \citenamefont {de~France}, \citenamefont {Gadea}, \citenamefont
  {Gonzalez}, \citenamefont {Gottardo}, \citenamefont {Harkness}, \citenamefont
  {Hess}, \citenamefont {Jungclaus}, \citenamefont {Ka\ifmmode
  \mbox{\c{s}}\else \c{s}\fi{}ka\ifmmode~\mbox{\c{s}}\else \c{s}\fi{}},
  \citenamefont {Korten}, \citenamefont {Lenzi}, \citenamefont {Leoni},
  \citenamefont {Ljungvall}, \citenamefont {Menegazzo}, \citenamefont
  {Mengoni}, \citenamefont {Million}, \citenamefont {Napoli}, \citenamefont
  {Nyberg}, \citenamefont {Podolyak}, \citenamefont {Pullia}, \citenamefont
  {Quintana~Arn\'es}, \citenamefont {Recchia}, \citenamefont {Redon},
  \citenamefont {Reiter}, \citenamefont {D.Salsac}, \citenamefont {Sanchis},
  \citenamefont {\ifmmode \mbox{\c{S}}\else
  \c{S}\fi{}enyi\ifmmode~\breve{g}\else \u{g}\fi{}it}, \citenamefont
  {Siciliano}, \citenamefont {Sohler}, \citenamefont {Stezowski}, \citenamefont
  {Theisen},\ and\ \citenamefont {Valiente~Dob\'on}}]{PhysRevC.107.014324}%
  \BibitemOpen
  \bibfield  {author} {\bibinfo {author} {\bibfnamefont {E.}~\bibnamefont
  {Cl\'ement}}, \bibinfo {author} {\bibfnamefont {A.}~\bibnamefont {Lemasson}},
  \bibinfo {author} {\bibfnamefont {M.}~\bibnamefont {Rejmund}}, \bibinfo
  {author} {\bibfnamefont {B.}~\bibnamefont {Jacquot}}, \bibinfo {author}
  {\bibfnamefont {D.}~\bibnamefont {Ralet}}, \bibinfo {author} {\bibfnamefont
  {C.}~\bibnamefont {Michelagnoli}}, \bibinfo {author} {\bibfnamefont
  {D.}~\bibnamefont {Barrientos}}, \bibinfo {author} {\bibfnamefont
  {P.}~\bibnamefont {Bednarczyk}}, \bibinfo {author} {\bibfnamefont
  {G.}~\bibnamefont {Benzoni}}, \bibinfo {author} {\bibfnamefont {A.~J.}\
  \bibnamefont {Boston}}, \bibinfo {author} {\bibfnamefont {A.}~\bibnamefont
  {Bracco}}, \bibinfo {author} {\bibfnamefont {B.}~\bibnamefont {Cederwall}},
  \bibinfo {author} {\bibfnamefont {M.}~\bibnamefont {Ciemala}}, \bibinfo
  {author} {\bibfnamefont {J.}~\bibnamefont {Collado}}, \bibinfo {author}
  {\bibfnamefont {F.}~\bibnamefont {Crespi}}, \bibinfo {author} {\bibfnamefont
  {C.}~\bibnamefont {Domingo-Pardo}}, \bibinfo {author} {\bibfnamefont
  {J.}~\bibnamefont {Dudouet}}, \bibinfo {author} {\bibfnamefont {H.~J.}\
  \bibnamefont {Eberth}}, \bibinfo {author} {\bibfnamefont {G.}~\bibnamefont
  {de~France}}, \bibinfo {author} {\bibfnamefont {A.}~\bibnamefont {Gadea}},
  \bibinfo {author} {\bibfnamefont {V.}~\bibnamefont {Gonzalez}}, \bibinfo
  {author} {\bibfnamefont {A.}~\bibnamefont {Gottardo}}, \bibinfo {author}
  {\bibfnamefont {L.}~\bibnamefont {Harkness}}, \bibinfo {author}
  {\bibfnamefont {H.}~\bibnamefont {Hess}}, \bibinfo {author} {\bibfnamefont
  {A.}~\bibnamefont {Jungclaus}}, \bibinfo {author} {\bibfnamefont
  {A.}~\bibnamefont {Ka\ifmmode \mbox{\c{s}}\else
  \c{s}\fi{}ka\ifmmode~\mbox{\c{s}}\else \c{s}\fi{}}}, \bibinfo {author}
  {\bibfnamefont {W.}~\bibnamefont {Korten}}, \bibinfo {author} {\bibfnamefont
  {S.~M.}\ \bibnamefont {Lenzi}}, \bibinfo {author} {\bibfnamefont
  {S.}~\bibnamefont {Leoni}}, \bibinfo {author} {\bibfnamefont
  {J.}~\bibnamefont {Ljungvall}}, \bibinfo {author} {\bibfnamefont
  {R.}~\bibnamefont {Menegazzo}}, \bibinfo {author} {\bibfnamefont
  {D.}~\bibnamefont {Mengoni}}, \bibinfo {author} {\bibfnamefont
  {B.}~\bibnamefont {Million}}, \bibinfo {author} {\bibfnamefont {D.~R.}\
  \bibnamefont {Napoli}}, \bibinfo {author} {\bibfnamefont {J.}~\bibnamefont
  {Nyberg}}, \bibinfo {author} {\bibfnamefont {Z.}~\bibnamefont {Podolyak}},
  \bibinfo {author} {\bibfnamefont {A.}~\bibnamefont {Pullia}}, \bibinfo
  {author} {\bibfnamefont {B.}~\bibnamefont {Quintana~Arn\'es}}, \bibinfo
  {author} {\bibfnamefont {F.}~\bibnamefont {Recchia}}, \bibinfo {author}
  {\bibfnamefont {N.}~\bibnamefont {Redon}}, \bibinfo {author} {\bibfnamefont
  {P.}~\bibnamefont {Reiter}}, \bibinfo {author} {\bibfnamefont
  {M.}~\bibnamefont {D.Salsac}}, \bibinfo {author} {\bibfnamefont
  {E.}~\bibnamefont {Sanchis}}, \bibinfo {author} {\bibfnamefont
  {M.}~\bibnamefont {\ifmmode \mbox{\c{S}}\else
  \c{S}\fi{}enyi\ifmmode~\breve{g}\else \u{g}\fi{}it}}, \bibinfo {author}
  {\bibfnamefont {M.}~\bibnamefont {Siciliano}}, \bibinfo {author}
  {\bibfnamefont {D.}~\bibnamefont {Sohler}}, \bibinfo {author} {\bibfnamefont
  {O.}~\bibnamefont {Stezowski}}, \bibinfo {author} {\bibfnamefont
  {C.}~\bibnamefont {Theisen}},\ and\ \bibinfo {author} {\bibfnamefont {J.~J.}\
  \bibnamefont {Valiente~Dob\'on}},\ }\bibfield  {title} {\bibinfo {title}
  {Spectroscopic quadrupole moments in $^{124}\mathrm{Xe}$},\ }\href
  {https://doi.org/10.1103/PhysRevC.107.014324} {\bibfield  {journal} {\bibinfo
   {journal} {Phys. Rev. C}\ }\textbf {\bibinfo {volume} {107}},\ \bibinfo
  {pages} {014324} (\bibinfo {year} {2023})}\BibitemShut {NoStop}%
\bibitem [{\citenamefont {Peters}\ \emph {et~al.}(2019)\citenamefont {Peters},
  \citenamefont {Stuchbery}, \citenamefont {Chakraborty}, \citenamefont
  {Crider}, \citenamefont {Ashley}, \citenamefont {Kumar}, \citenamefont
  {McEllistrem}, \citenamefont {Prados-Est\'evez},\ and\ \citenamefont
  {Yates}}]{PhysRevC.99.064321}%
  \BibitemOpen
  \bibfield  {author} {\bibinfo {author} {\bibfnamefont {E.~E.}\ \bibnamefont
  {Peters}}, \bibinfo {author} {\bibfnamefont {A.~E.}\ \bibnamefont
  {Stuchbery}}, \bibinfo {author} {\bibfnamefont {A.}~\bibnamefont
  {Chakraborty}}, \bibinfo {author} {\bibfnamefont {B.~P.}\ \bibnamefont
  {Crider}}, \bibinfo {author} {\bibfnamefont {S.~F.}\ \bibnamefont {Ashley}},
  \bibinfo {author} {\bibfnamefont {A.}~\bibnamefont {Kumar}}, \bibinfo
  {author} {\bibfnamefont {M.~T.}\ \bibnamefont {McEllistrem}}, \bibinfo
  {author} {\bibfnamefont {F.~M.}\ \bibnamefont {Prados-Est\'evez}},\ and\
  \bibinfo {author} {\bibfnamefont {S.~W.}\ \bibnamefont {Yates}},\ }\bibfield
  {title} {\bibinfo {title} {Emerging collectivity from the nuclear structure
  of $^{132}\mathrm{Xe}$: Inelastic neutron scattering studies and shell-model
  calculations},\ }\href {https://doi.org/10.1103/PhysRevC.99.064321}
  {\bibfield  {journal} {\bibinfo  {journal} {Phys. Rev. C}\ }\textbf {\bibinfo
  {volume} {99}},\ \bibinfo {pages} {064321} (\bibinfo {year}
  {2019})}\BibitemShut {NoStop}%
\bibitem [{\citenamefont {Urban}\ \emph {et~al.}(2019)\citenamefont {Urban},
  \citenamefont {Rz\c{a}ca-Urban}, \citenamefont {Wi\ifmmode~\acute{s}\else
  \'{s}\fi{}niewski}, \citenamefont {Smith}, \citenamefont {Simpson},\ and\
  \citenamefont {Ahmad}}]{PhysRevC.100.014319}%
  \BibitemOpen
  \bibfield  {author} {\bibinfo {author} {\bibfnamefont {W.}~\bibnamefont
  {Urban}}, \bibinfo {author} {\bibfnamefont {T.}~\bibnamefont
  {Rz\c{a}ca-Urban}}, \bibinfo {author} {\bibfnamefont {J.}~\bibnamefont
  {Wi\ifmmode~\acute{s}\else \'{s}\fi{}niewski}}, \bibinfo {author}
  {\bibfnamefont {A.~G.}\ \bibnamefont {Smith}}, \bibinfo {author}
  {\bibfnamefont {G.~S.}\ \bibnamefont {Simpson}},\ and\ \bibinfo {author}
  {\bibfnamefont {I.}~\bibnamefont {Ahmad}},\ }\bibfield  {title} {\bibinfo
  {title} {First observation of $\ensuremath{\gamma}$-soft and triaxial bands
  in $\mathrm{Zr}$ isotopes},\ }\href
  {https://doi.org/10.1103/PhysRevC.100.014319} {\bibfield  {journal} {\bibinfo
   {journal} {Phys. Rev. C}\ }\textbf {\bibinfo {volume} {100}},\ \bibinfo
  {pages} {014319} (\bibinfo {year} {2019})}\BibitemShut {NoStop}%
\bibitem [{\citenamefont {Karayonchev}\ \emph {et~al.}(2020)\citenamefont
  {Karayonchev}, \citenamefont {Jolie}, \citenamefont {Blazhev}, \citenamefont
  {Dewald}, \citenamefont {Esmaylzadeh}, \citenamefont {Fransen}, \citenamefont
  {H\"afner}, \citenamefont {Knafla}, \citenamefont {Litzinger}, \citenamefont
  {M\"uller-Gatermann}, \citenamefont {R\'egis}, \citenamefont {Schomacker},
  \citenamefont {Vogt}, \citenamefont {Warr}, \citenamefont {Leviatan},\ and\
  \citenamefont {Gavrielov}}]{PhysRevC.102.064314}%
  \BibitemOpen
  \bibfield  {author} {\bibinfo {author} {\bibfnamefont {V.}~\bibnamefont
  {Karayonchev}}, \bibinfo {author} {\bibfnamefont {J.}~\bibnamefont {Jolie}},
  \bibinfo {author} {\bibfnamefont {A.}~\bibnamefont {Blazhev}}, \bibinfo
  {author} {\bibfnamefont {A.}~\bibnamefont {Dewald}}, \bibinfo {author}
  {\bibfnamefont {A.}~\bibnamefont {Esmaylzadeh}}, \bibinfo {author}
  {\bibfnamefont {C.}~\bibnamefont {Fransen}}, \bibinfo {author} {\bibfnamefont
  {G.}~\bibnamefont {H\"afner}}, \bibinfo {author} {\bibfnamefont
  {L.}~\bibnamefont {Knafla}}, \bibinfo {author} {\bibfnamefont
  {J.}~\bibnamefont {Litzinger}}, \bibinfo {author} {\bibfnamefont
  {C.}~\bibnamefont {M\"uller-Gatermann}}, \bibinfo {author} {\bibfnamefont
  {J.-M.}\ \bibnamefont {R\'egis}}, \bibinfo {author} {\bibfnamefont
  {K.}~\bibnamefont {Schomacker}}, \bibinfo {author} {\bibfnamefont
  {A.}~\bibnamefont {Vogt}}, \bibinfo {author} {\bibfnamefont {N.}~\bibnamefont
  {Warr}}, \bibinfo {author} {\bibfnamefont {A.}~\bibnamefont {Leviatan}},\
  and\ \bibinfo {author} {\bibfnamefont {N.}~\bibnamefont {Gavrielov}},\
  }\bibfield  {title} {\bibinfo {title} {Tests of collectivity in
  $^{98}\mathrm{Zr}$ by absolute transition rates},\ }\href
  {https://doi.org/10.1103/PhysRevC.102.064314} {\bibfield  {journal} {\bibinfo
   {journal} {Phys. Rev. C}\ }\textbf {\bibinfo {volume} {102}},\ \bibinfo
  {pages} {064314} (\bibinfo {year} {2020})}\BibitemShut {NoStop}%
\end{thebibliography}%

\end{document}